\documentclass[
reprint,
amsmath,
amssymb,
aps,
showkeys,
]{revtex4-2}
\usepackage[displaymath, mathlines]{lineno}
\usepackage{amsmath}
\usepackage{amssymb}
\usepackage{array}	
\usepackage{graphicx}
\usepackage{siunitx}
\usepackage{booktabs}
\usepackage[english,ngerman]{babel}
\usepackage[table]{xcolor}
\usepackage{nicematrix}
\RequirePackage{float}
\floatstyle{plain}
\usepackage[hang]{subfigure}
\usepackage[utf8]{inputenc}

\addto\captionsngerman{

}
\usepackage[section]{placeins}

\newcommand{\ra}[1]{\renewcommand{\arraystretch}{#1}}

\begin{document}
\author{Martin He\ss ler}
\email{m\_{}hess23@uni-muenster.de}
\affiliation{Institute for Theoretical Physics, WWU M\"unster, M\"unster Germany\\}
 \altaffiliation[Also at ]{Center for Nonlinear Science, WWU M\"unster, M\"unster Germany\\}

\author{Oliver Kamps}
 \email{okamp@uni-muenster.de}
\affiliation{Center for Nonlinear Science, WWU M\"unster, M\"unster Germany\\}

\title{Bayesian on-line anticipation of critical transitions}

\date{\today}

\begin{abstract}
\linenumbers
The design of reliable indicators to anticipate critical transitions in complex systems is an important task in order to detect a coming sudden  regime shift and to take action in order to either prevent it or mitigate its consequences. We present a data-driven method based on the estimation of a parameterized nonlinear stochastic differential equation that allows for a robust anticipation of critical transitions even in the presence of strong noise levels like they are present in many real world systems. Since the parameter estimation is done by a Markov Chain Monte Carlo approach we have access to credibility bands allowing for a better interpretation of the reliability of the results. By introducing a Bayesian linear segment fit it is possible to give an estimate for the time horizon in which the transition will probably occur based on the current state of information. This approach is also able to handle nonlinear time dependencies of the parameter controlling the transition. In general the method could be used as a tool for on-line analysis to detect changes in the resilience of the system and to provide information on the probability of the occurrence of a critical transition in future.
\end{abstract}


\maketitle

\linenumbers
In complex systems in nature, technology and society smooth parameter changes can lead to sudden transitions between system states with very different behavior \cite{Scheffer2009book}. Examples for such critical transitions occurring at tipping points are climate changes like desertification \cite{a:Corrado2014}, greenhouse-icehouse transitions in the past \cite{a:livina2010, a:Livina2013, a:Livina2015,a:thompson2011, a:Lenton2012}, extinction of animal or bacteria species \cite{a:veraart11, a:scheffer09,a:dakos08, a:Dakos2014, a:Dakos2017}, allograft rejection \cite{a:izrailtyan}, chronic diseases \cite{a:Rikkert2016}, power outages \cite{a:cotilla-sanchez12, a:Ren2015}, the Laser threshold and the onset of convection \cite{Haken2004}, transitions in human movement \cite{Haken1985biolcyber} and much more \cite{b:strogatz, a:jurczyk2017, a:Leemput2013, Chadefaux2014,a:Gopalakrishnan2016,a:Lim2011}. Sometimes, these transitions have severe consequences \cite{a:Boers2021, a:Klose2020, a:Ritchie2021, a:Lohmann2021, c:Lenton2019} and it would be important to find methods to anticipate critical transitions and strategies to prevent or avoid damage \cite{a:Batt2019, a:Hagstrom2017,a:Ewel2001}. 

In many natural systems where critical transitions occur, the detailed mechanisms are barely understood. Nevertheless, it is well known that in many cases complex systems composed out of a large number of nonlinear interacting parts can be described by only a few macroscopic degrees of freedom if they are close to a critical transition \cite{Haken2004}. The critical transition of the complex system connected to a smooth parameter change corresponds to a bifurcation of the evolution equation. This bifurcation induced scenario is sometimes called \textit{B-tipping} \cite{a:Ritchie2017, a:Ashwin2012}. Since these bifurcation scenarios are rather universal one can expect that for a large class of systems, like the examples mentioned above, a coming bifurcation can be anticipated by observing universal indicators that are correlated to the loss of linear stability or by directly observing an estimate of a numerical linear stability analysis. 

So in recent years a plenty of time series analysis methods based on the assumption of universal indicators have been developed to solve the problem of anticipating a coming bifurcation  \cite{a:dakos12,a:Dakos2009, a:Liang2017, a:Scheffer2015}. A non-parametric ansatz is implemented in \cite{a:carpenter11, a:Anvari2016} in which the estimation of a drift-diffusion-jump model is suggested in case of high-frequency time series. The most simple \textit{leading indicators} to anticipate a critical transition are the autocorrelation at lag-1 combined with the standard deviation based on the phenomenon of \textit{critical slowing down} prior to a bifurcation scenario, the kurtosis based on the changing shape of the underlying potential and the skewness based on \textit{flickering} in noisy systems \cite{a:scheffer12, a:Xie2018}. 
The measures are evaluated in rolling windows and need a preparation of the data, e.g. detrending in each window. They are loosely connected to the mathematical description of complex systems and their positive trends are sometimes misleading or difficult to interpret, because of missing uncertainty estimates and sometimes autocorrelation and standard deviation do not increase at the same time as they should if they are candidates for an uprising transition \cite{a:Boerlijst2013, a:Gsell2016, a:Ditlevsen2010}.
In general, the briefly mentioned stability measures and leading indicators suffer from their difficult interpretation due to noisy trends and missing uncertainty estimates. Besides, selecting historical test data based on the knowledge that those datasets exhibit a transition can lead to statistical mistakes in the leading indicators' performance and quality \cite{a:Boettiger2012}. Furthermore, they are rather susceptible for artefacts that are not related to critical transitions, but caused by a changing variance, sub-optimal detrending of the time series or simply preprocessing issues \cite{a:Wilkat2019,a:Clements2015,a:Hastings2010,a:Perretti2012,a:Dablander2020}. Additionally, high noise levels can have strong negative influence on the performance of leading indicators \cite{a:Perretti2012}. Another important issue is the estimation of the point in time at which the transition will probably take place. Even in cases where a reliable leading indicator can be computed it is still difficult to give an estimate for the time horizon in which the transition will occur. Though it is possible to extrapolate a straight line based on the most current calculated leading indicator values, this strategy will fail to give an reliable estimate in more complicated situations, e.g. the onset of a parameter trend of unknown shape out of a regime with a constant control parameter. 

In the following we will develop a method to anticipate critical transitions in complex systems which solves many of the shortcomings of the aforementioned approaches. The method is based on the assumption that the measured time series can be modeled by a nonlinear stochastic differential equation (SDE). Therefore, the parameters of the model can be linked directly to linear stability analysis and the inclusion of higher order terms in the ansatz for the SDE allows for a reliable statement about the stability of the system even in the presence of strong noise. The parameters' estimation is done in a Bayesian way providing access to the complete posterior distribution of the parameters and subsequently allowing for the specification of credibility intervals.  To estimate the beginning of the transition towards a different regime and the time horizon in which the transition can be expected we introduce a Bayesian non-parametric linear segment fit \cite{b:linden2014, a:dose2004} which is able to react to sudden changes of the control parameter trend under rather simple assumptions. 

The paper is divided into three parts. In section \ref{sec: How to model} the model assumptions and the estimation procedure are introduced. In section \ref{sec: results} the method is applied as an example to two synthetic models and a statistical comparison to the above mentioned leading indicators autocorrelation, standard deviation, kurtosis and skewness is given. The method is also applied to a non-Markovian dataset to get an impression of the reliability under imperfect circumstances and the non-parametric linear segment fit approach is presented as an example. At last, a short summary and conclusion is presented in section \ref{sec: conclusion}.\\ 

\section{\label{sec: How to model} Developing the method}

\subsection{\label{subsec: Model approach} The model approach for the observed data}

The system under consideration is assumed to be relaxed onto a stable fixed point. It explores the phase space close to the fixed point due to noise and as a result one observes a fluctuating time series of the system. The starting hypothesis is similar to the ansatz in \cite{a:carpenter11} and assumes that the observed signal can be described by a SDE of the Langevin \cite{b:Kloeden1992} form

\begin{linenomath*}
\begin{equation}
\dot{x}(x,t) = h(x(t),t) + g(x(t), t) \Gamma (t).\label{eq:langevin}
\end{equation}
\end{linenomath*}
where $h(x,t)$ is the nonlinear deterministic part, the so-called drift, and $g(x(t), t)$ is the diffusion term. The noise $\Gamma (t)$ is assumed to be Gaussian and $\delta$-correlated. For the remainder of the article we assume that we can make the approximation that the noise strength does not significantly depend on the state $x(t)$ an we can set $g(x(t), t) = const. = \sigma$. \\
The loss of stability of the fixed point followed by a bifurcation is described by the change of the sign of the slope 
\begin{linenomath*}
\begin{equation}
\zeta = \left.\frac{\text{d}h(x)}{\text{d}x}\right\vert_{x = x^*} 
\end{equation}
\end{linenomath*}
of the nonlinear drift at the fixed point $x^*$. Since we assume to be in a fixed point and close to a bifurcation we develop $h(x,t)$ into a Taylor series up to order three which is sufficient to describe the normal forms of simple bifurcation scenarios \cite{b:strogatz}. This results in
\begin{linenomath*}
\begin{equation}
\begin{split}
h(x(t),t) &= \alpha_0(t) + \alpha_1(t) (x - x^*) + \alpha_2(t) (x - x^*)^2 \\
&+ \alpha_3(t) (x - x^*)^3 + \mathcal{O}(x^4)  \label{eq:taylor} 
\end{split},
\end{equation}
\end{linenomath*}
so that the information on the linear stability is incorporated in $\alpha_1$. For practical reasons equation \ref{eq:taylor} is used in the form
\begin{linenomath*}
\begin{equation}
\begin{split}
    h_{\rm MC}(x(t),t) &= \theta_0 (t; x^*) + \theta_1 (t; x^*) \cdot x + \theta_2 (t; x^*) \cdot x^2 \\ 
    &+ \theta_3 (t; x^*) \cdot x^3 + \mathcal{O}(x^4)
    \end{split}
\end{equation}
\end{linenomath*}
in the numerical approach, where an arbitrary fixed point $x^*$ is incorporated in the coefficients $\underline{\theta}$ by algebraic transformation and comparison of coefficients. The polynomial model up to order three is denoted as $\rho (x)$, a simple linear model is called $\lambda (x)$. Now, the time series can be divided into short windows where the parameters can be assumed to be nearly constant. Estimating the parameters within the windows and observing their change from window to window give us information on the evolution of the stability of the system in time. Since one is interested only in the change of the slope at the fixed point one could restrict this ansatz to the first order term $\alpha_1$. But if the noise is strong enough to drive the system away from the fixed point, like it is assumed to be e.g. in ecological systems \cite{a:Perretti2012}, contributions of the higher order terms may become significant and their incorporation leads to a much more reliable estimation of the slope $\zeta$. Additionally, close to the bifurcation the nonlinear terms play a more significant role since the first order term goes to zero which is the reason for critical slowing down and critical fluctuations.\\ 

\subsection{\label{sec: How to Bayes} Bayesian estimation of the model parameters}

We have to estimate the model parameters from the data to observe the change of the stability within the moving windows. Here we choose a Bayesian approach since this allows for the computation of the full posterior distribution of the parameters via Markov Chain Monte Carlo (MCMC) and a consistent inclusion of prior information \cite{b:linden2014}. Using Bayes' theorem the posterior distribution of the parameters is defined as 
\begin{linenomath*}
\begin{equation}
p(\underline{\theta},\sigma |\underline{d}, \mathcal{I}) = \frac{p(\underline{d}|\underline{\theta}, \sigma, \mathcal{I}) \cdot p(\underline{\theta}, \sigma |\mathcal{I})}{p(\underline{d}  | \mathcal{I})}
\end{equation}
\end{linenomath*}
with the evidence  $p(\underline{d} | \mathcal{I})$, which normalizes the posterior distribution, the likelihood $p(\underline{d}|\underline{\theta}, \mathcal{I})$ and the prior distribution $p(\underline{\theta}|\mathcal{I})$. The posterior allows for the consistent definition of credibility intervals of the estimated parameters.
The likelihood $p(\underline{d}|\underline{\theta}, \mathcal{I})$ relates the measured data $\underline{d}$ to the underlying model and is given by the transition probability of the Markov process defined by \eqref{eq:langevin}. The transition probability  $p(x, t + \tau | x', t)$ for subsequent times $t$ and $t'$ with $\tau = t - t' \longrightarrow 0$ which is also denoted as short time propagator \cite{Risken96} is defined as 
\begin{linenomath*}
\begin{equation}
p(x, t + \tau | x', t) = \frac{1}{\sqrt{2\pi \sigma^2\tau }}  \exp\left(-\frac{[x - x' - h_{\rm MC}(x',t)\tau]^2}{2 \sigma^2\tau}\right).
\end{equation}
\end{linenomath*}
The priors are generally chosen in a broad range to allow for the determination of the parameters due to the available data instead of strong prior assumptions. For the linear parts of the deterministic functions $\lambda (x)$ and $\rho (x)$ the uninformative Jeffreys' prior \cite{b:linden2014}
\begin{linenomath*}
\begin{equation}
p_{\rm prior}(\theta_0,\theta_1) = \frac{1}{2\pi (1+\theta_1^2)^\frac{3}{2}}
\end{equation}
\end{linenomath*}
is chosen. It is improper in the constant $\theta_0$, but that does not matter for parameter estimation. The scale variable $\sigma$ is modelled by Jeffreys' scale prior \cite{b:linden2014}
\begin{linenomath*}
\begin{equation}
p_{\rm prior}(\sigma ) = \frac{1}{\sigma}.
\end{equation}
\end{linenomath*}
These choices simulate best the real situation in which no or just poor prior information is available.
Furthermore, it is taken care that the parameters of the higher orders are initially able to contribute in a similar magnitude to the deterministic dynamics as the linear ones by wide Gaussian priors
\begin{linenomath*}
\begin{equation}
\begin{aligned}
p_{\rm prior}(\theta_2) &= \mathcal{N}(\mu, \sigma_{\theta_2}),  \\
p_ {\rm prior}(\theta_3) &= \mathcal{N}(\mu, \sigma_{\theta_3})
\end{aligned}
\end{equation}
\end{linenomath*}
centred around the mean $\mu = 0$ with standard deviations $\sigma_{\theta_i}$ in an adequate range.\\  
The posterior can be computed approximately by hand for the simple linear approximation $\lambda (x)$. The detailed solution involves the assumption of a Gaussian posterior, the use of perturbation theory and a Taylor expansion. It can be found in \cite{b:linden2014}. However, the posterior distribution for more complicated models as the third order polynom $\rho (x)$ cannot be computed analytically, but by using a MCMC approach.  Here we use an affine-invariant ensemble sampler with a so-called \textit{stretch step}. It implements an ensemble $C = \lbrace X_k \rbrace$ of $k$ Markov chains $X_k$ at the same time. The proposal for the Markov chain $X_i$ is drawn from a proposal distribution that is based on the current states of the complementary ensemble $C_{[i]} = \lbrace X_j, \forall j \neq i\rbrace$. This shortens the integrated autocorrelation length and improves the computational performance. We use the version implemented in the python package \textit{emcee} \cite{a:Foreman-Mackey2013}. 

Based on the posterior distribution $p(\underline{\theta},\sigma |\underline{d}, \mathcal{I})$ we can compute the posterior pdfs of the slope by marginalization. The posterior pdf of the slope $\theta_1$ of the linear model $\zeta=\theta_1$ is easily accessible via
\begin{linenomath*}  
\begin{equation}
    p(\zeta|\underline{d},\mathcal{I}) = \int  \!\!  p(\underline{\theta},\sigma |\underline{d},\mathcal{I}) \text{d}\theta_0 \text{d}\sigma .  
\end{equation}
\end{linenomath*}
and the posterior pdf of the slope of the third order drift model $\rho (x)$ is given by 
\begin{linenomath*}
\begin{equation}
    p(\zeta |\underline{d},\mathcal{I}) = \int \!\! p(\underline{\theta},\sigma |\underline{d},\mathcal{I}) \delta\left( \zeta - \left.\frac{\text{d}h(x)}{\text{d}x}\right\vert_{x = x^*}\right) \text{d}\underline{\theta} \text{d}\sigma .
\end{equation}
\end{linenomath*}
Now we can define the credibility intervals as the $\SI{1}{\percent}$ to $\SI{99}{\percent}$- and $\SI{16}{\percent}$ to $\SI{84}{\percent}$-percentiles of the corresponding pdf. It is computed from a kernel density estimate of the corresponding pdf \cite{a:sklearn}.

\subsection{\label{subsec: Change points} Bayesian estimation of change points}

By using the method described in the last section we can reliably estimate the parameters of the model \eqref{eq:langevin} in a moving window. By observing the change of $\zeta$ from window to window we can detect the change in the degree of stability or more generally, the \textit{resilience} \cite{a:Scheffer2018} of the system, and can decide whether the system drifts towards an upcoming critical transition. Nevertheless, the method has to be extended to tackle two additional questions. In many situations the system is in stable operation and at a certain point in time it starts to change its stability for example due to the change of external conditions. Observing the fluctuations of the frequency in a power grid to anticipate an outage would be an example for this situation \cite{a:cotilla-sanchez12}. In this cases it would be important, e.g. for an on-line warning system, to detect the point where the system starts loosing its stability as soon as possible. The second point is the question at which time the transition will occur. In principle, the question can be answered by extrapolating the indicator values computed in the time windows by a straight line to find the point in time where the indicator crosses the critical value. Nevertheless, this approach is obviously not very reliable.
To solve the two problems we extend the method by including a non-parametric Bayesian linear segment fit into our method. Such a fitting approach is even able to be adapted in an on-line application and provides forecasts for possible critical transitions with corresponding uncertainties based on the current information. 
In general, the method assumes that the dataset can be fitted by a predefined number of linear segments divided by so-called \textit{change points} where the segments can change their slope. In case of one change point the dataset is fitted by two linear segments with different slopes connected at the change point. In our application the dataset is the set of slope indicator values $\underline{\zeta}$. If we assume $n$ change points we have $n+1$ linear segments 
\begin{linenomath*}
\begin{equation}
\phi_K (t|\underline{E}, \underline{\zeta}_{\rm ord}, \mathcal{I}) = \zeta_K \frac{t_{K+1}-t}{t_{K+1} - t_K} + \zeta_{K+1}\frac{t-t_K}{t_{K+1}-t_K},
\end{equation}  
\end{linenomath*}
for $t_K\leqslant t \leqslant t_{K+1}$ and the change point vector $\underline{E}$ with the entries $E_K = t_{K+1}$ for $K = 1,2,...,N-2$, whereby $N$ denotes the data size and the vector $\underline{\zeta}_{\rm ord}$ contains the design ordinates $\zeta_K$ at the assumed change point positions. Now, the goal is to calculate the probability 
\begin{linenomath*}
\begin{equation}
    p(\underline{E}|\underline{\zeta},\underline{t},\mathcal{I}) = \frac{p(\underline{E} \mid \underline{t}, \mathcal{I}) \cdot p(\underline{\zeta} \mid \underline{t}, \underline{E}, \mathcal{I})}{p(\underline{\zeta} \mid \underline{t}, \mathcal{I})}  \label{eq:changePointPosterior}
\end{equation}
\end{linenomath*}
that a certain configuration of change points is realised in the dataset $\underline{\zeta}$. The dataset $\underline{\zeta}$ is expanded by each step of the moving window approach. Furthermore, we assume that the system initially is in a steady state and that the slope calculated in the moving window fluctuates around a fixed value. In this case there is no change point within the actual dataset and the result is a flat distribution \eqref{eq:changePointPosterior}. If time proceeds a control parameter starts to change and causes the stability of the system to degrade. In consequence the slope indicator $\zeta$ will start to change and \eqref{eq:changePointPosterior} will be peaked around a certain $t_K$ which can then be identified as pint where the system starts to change its stability. In consequence, the data will be fitted by a nearly constant segment and a segment with finite slope beginning at the change point and we are able to detect the beginning destabilizing process. Given this information we can forecast the point where $\zeta$ crosses zero by computing the sum of the forecasts gained from all possible change point configurations weighted with their posterior probabilities. In the case of only one change point this would correspond to a simple straight line extrapolation of the data. However, it is not guaranteed that the time evolution of the control parameter is linear. The approach can be extended in a straightforward way to deal with a nonlinear dependence of the control parameter by introducing additional change points in the fit. This becomes more and more computational expensive so we restrict the ansatz in this paper to two change points. The details to calculate the posterior \eqref{eq:changePointPosterior} and the forecast can be found in \cite{b:linden2014, a:dose2004}.\\
Given the procedure above, the approach is able to react to sudden changes of the trend of the control parameter and enables us to extrapolate the trend based on the available data to find possible transition times $t_{\rm crit}$ including their uncertainty based on the model assumptions. The results of the described slope estimation procedure and an exemplary discussion of the Bayesian non-parametric transition anticipation method are presented in the following section \ref{sec: results}.

\section{\label{sec: results} Numerical results} 

In subsection \ref{subsec: Bayesian case studies} the Bayesian stability estimation approach is tested against two simulated datasets of which one passes through a fold and the other through a pitchfork bifurcation. A comparison to known leading indicator candidates is given in subsection \ref{subsec: comparison}. Moreover, the method is applied to a non-Markovian dataset in \ref{subsec: non markovian} to provide an example for the robustness of the evaluation under imperfect conditions. \\
Finally, in subsection \ref{subsec: prediction} the before mentioned non-parametric linear segment fit is applied as an example to a fold bifurcation model.\\
\subsection{\label{subsec: Bayesian case studies}Bayesian case studies with simulated data}
\begin{figure*}[!htbp]
\subfigure{
\label{subfig: fold data}
\includegraphics[width=0.32 \textwidth]{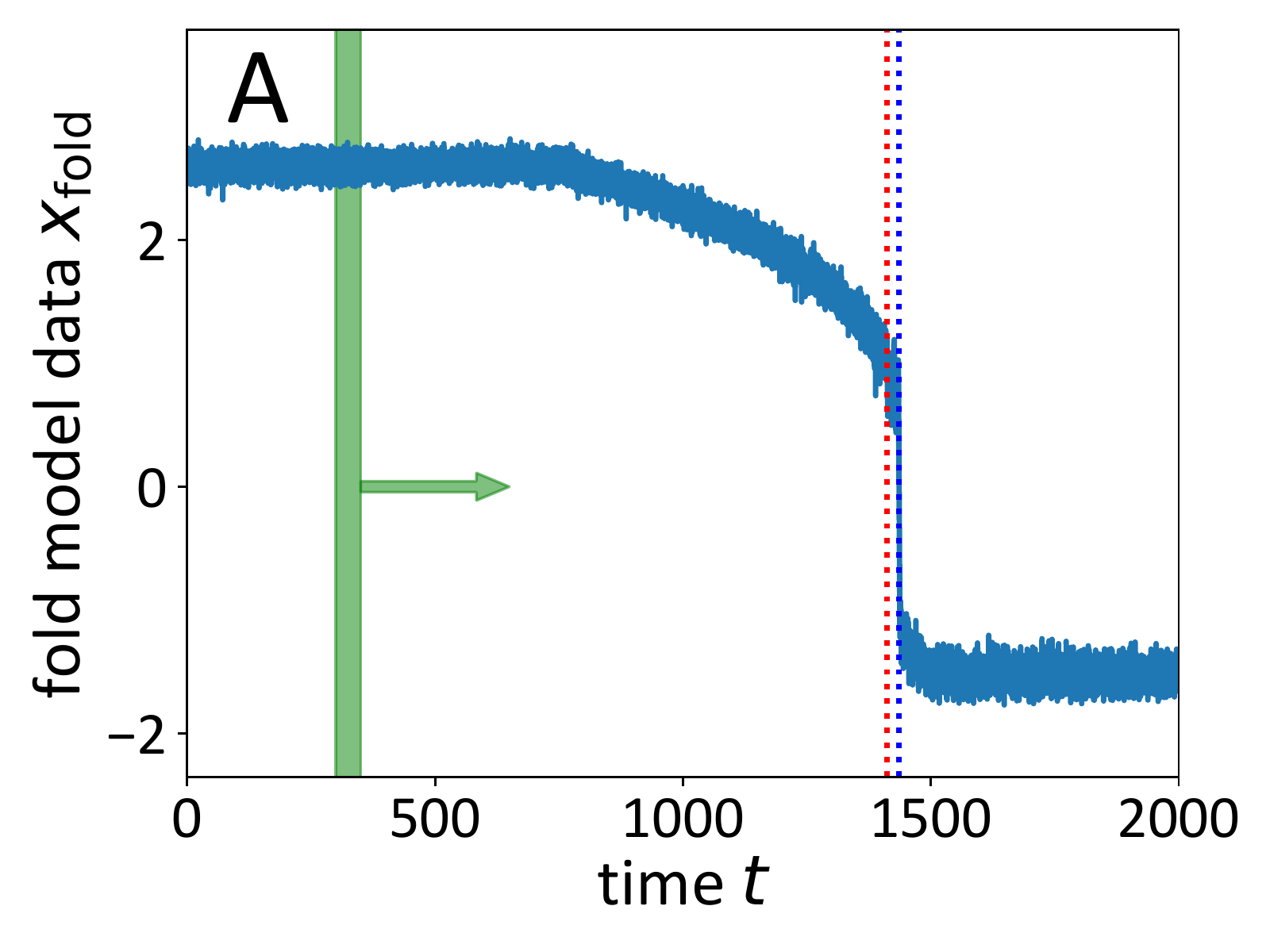}}
\subfigure{
\label{subfig: r param}
\includegraphics[width=.32 \textwidth]{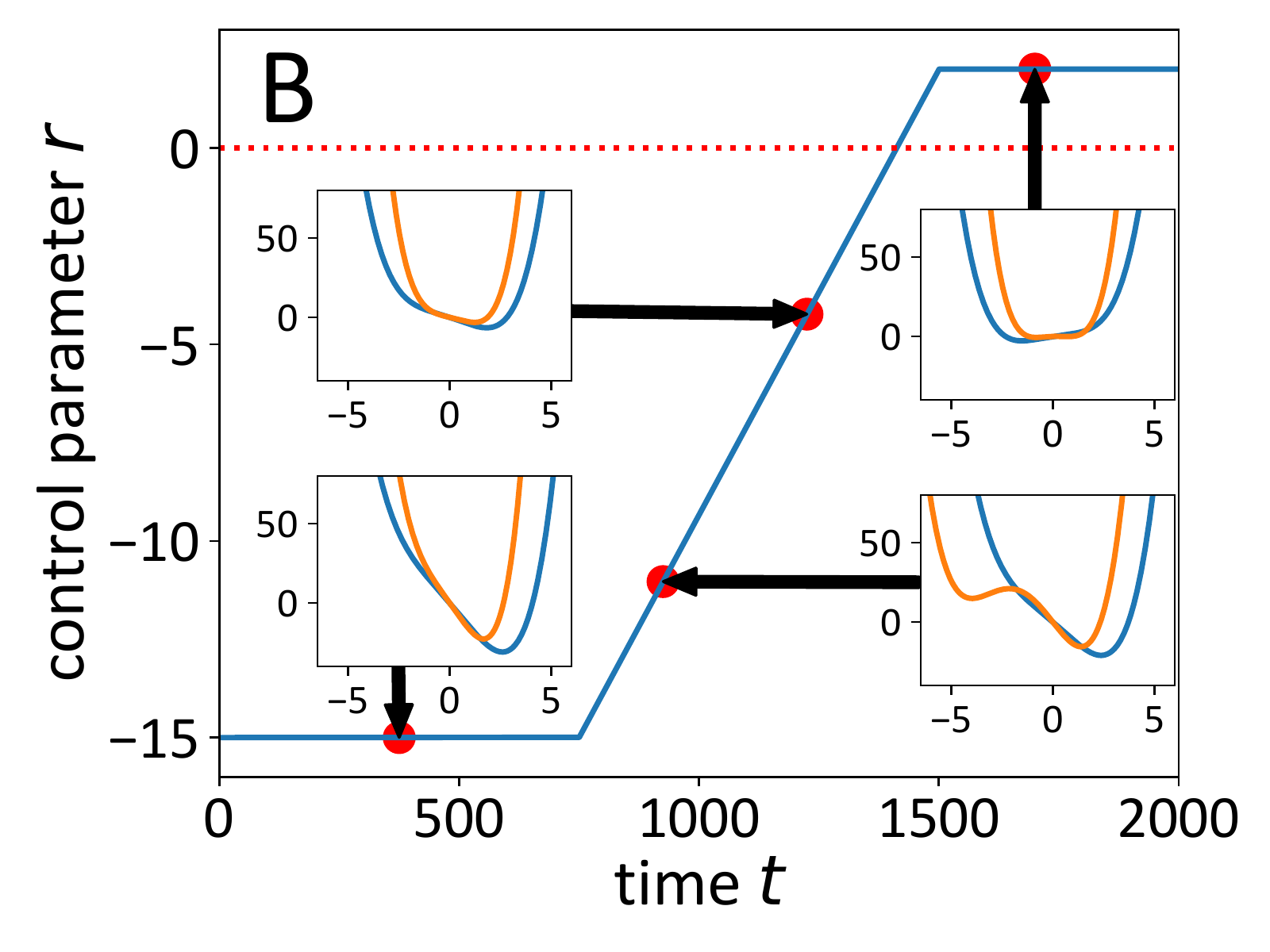}}
\subfigure{
\label{subfig: slope pdfs}
\includegraphics[width=.32 \textwidth]{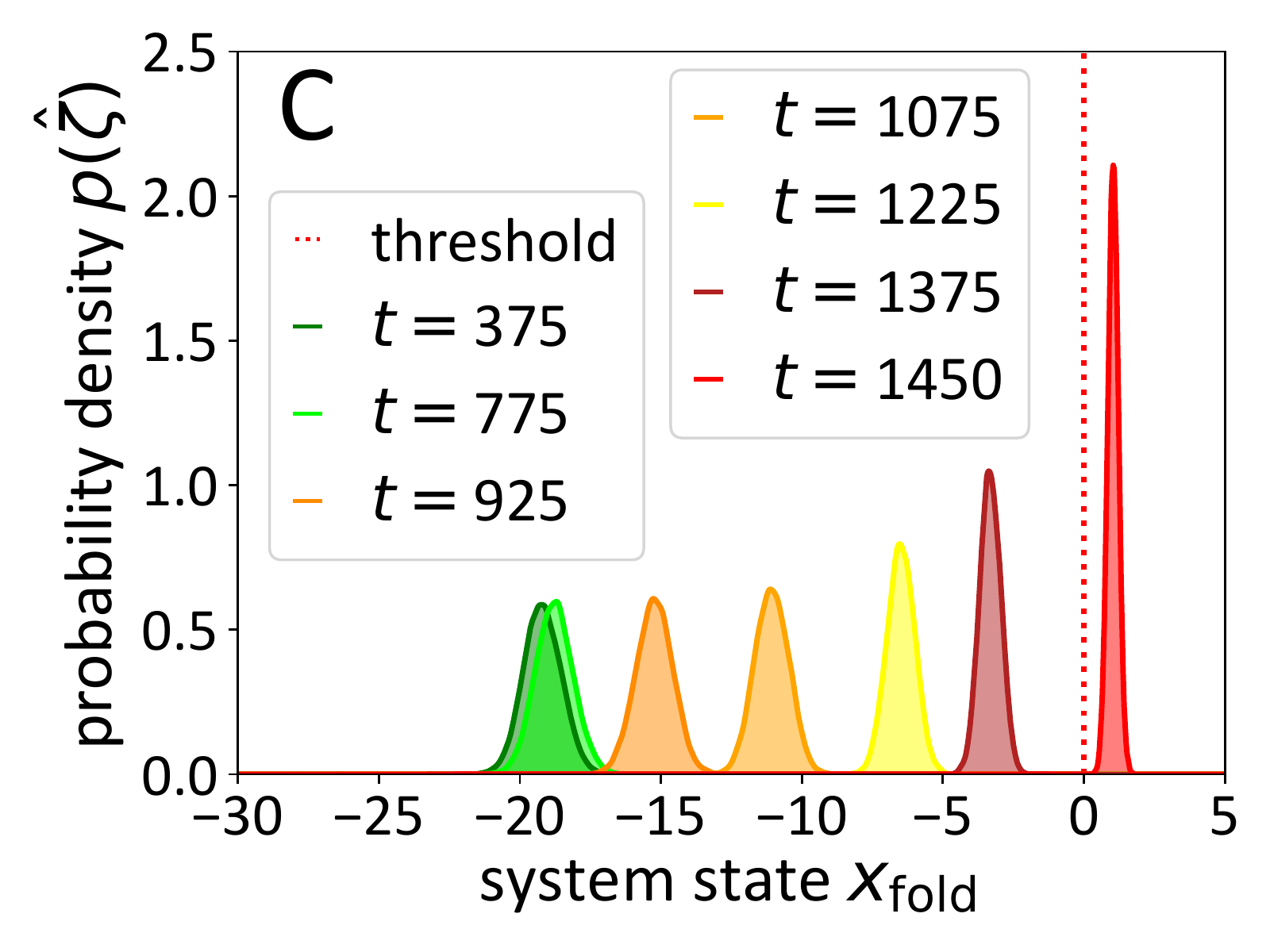}}
\subfigure{
\label{subfig: param01}
\includegraphics[width=0.32 \textwidth]{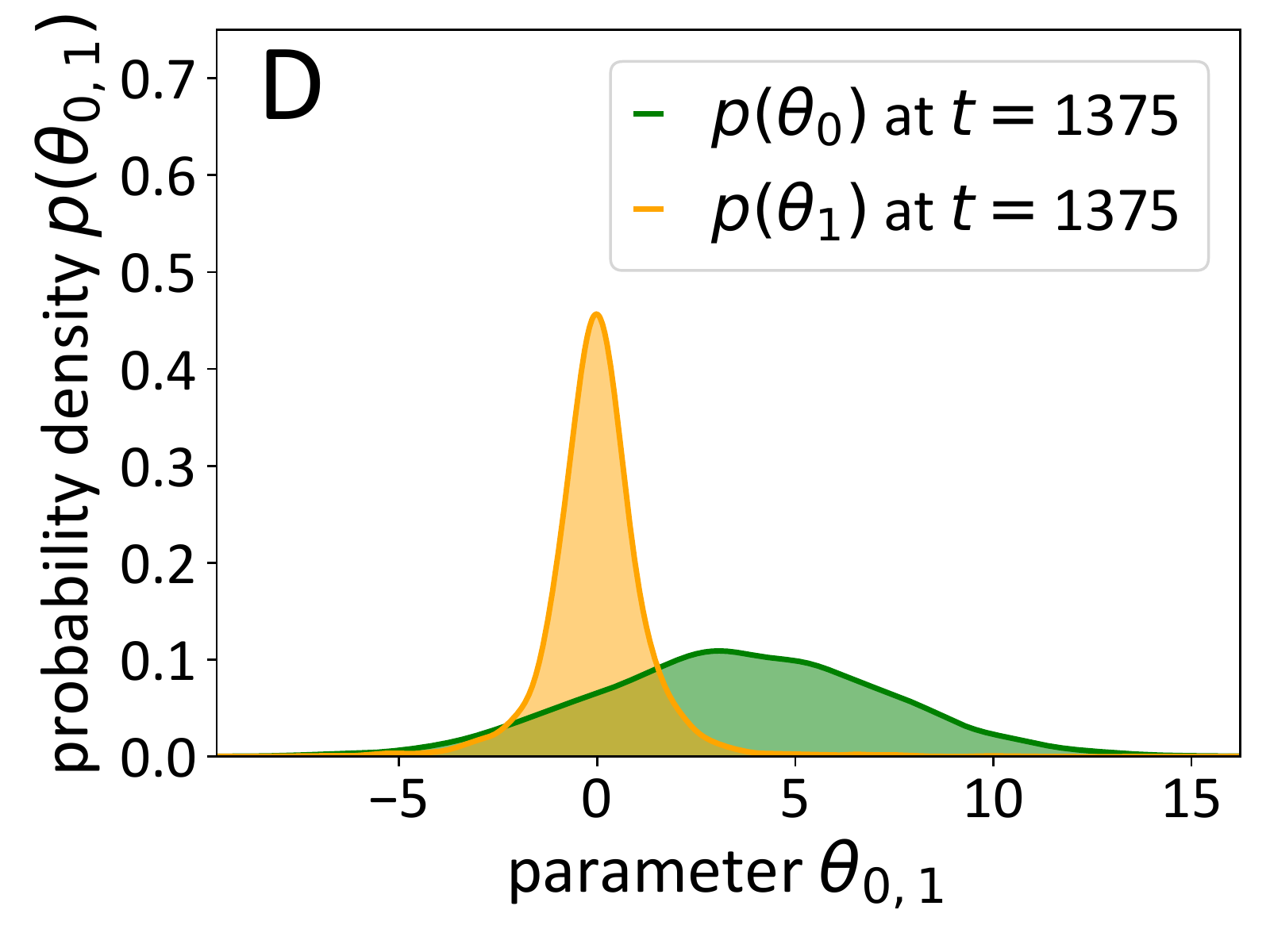}}
\subfigure{
\label{subfig: param23}
\includegraphics[width=.32 \textwidth]{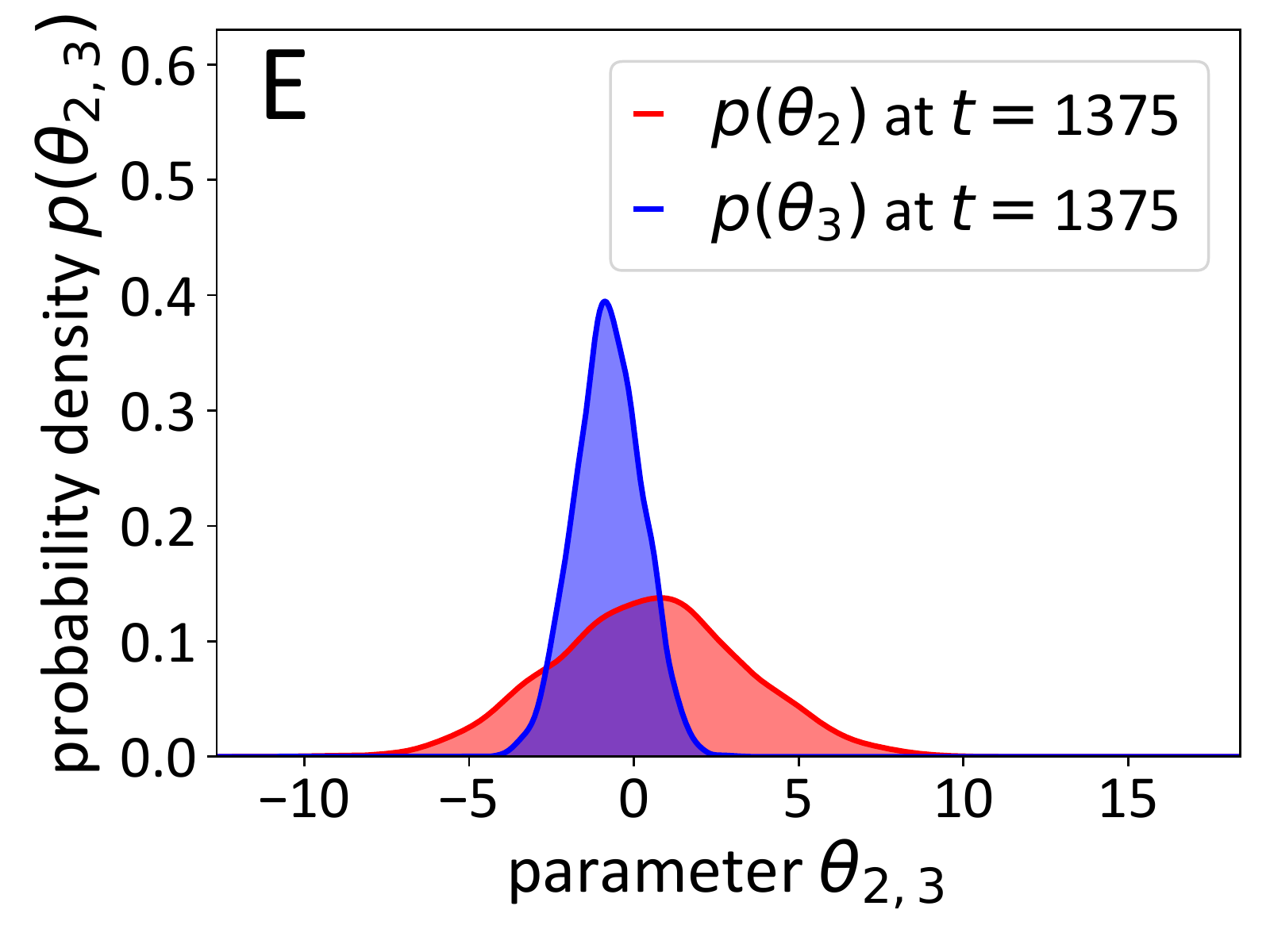}}
\subfigure{
\label{subfig: param4}
\includegraphics[width=.32 \textwidth]{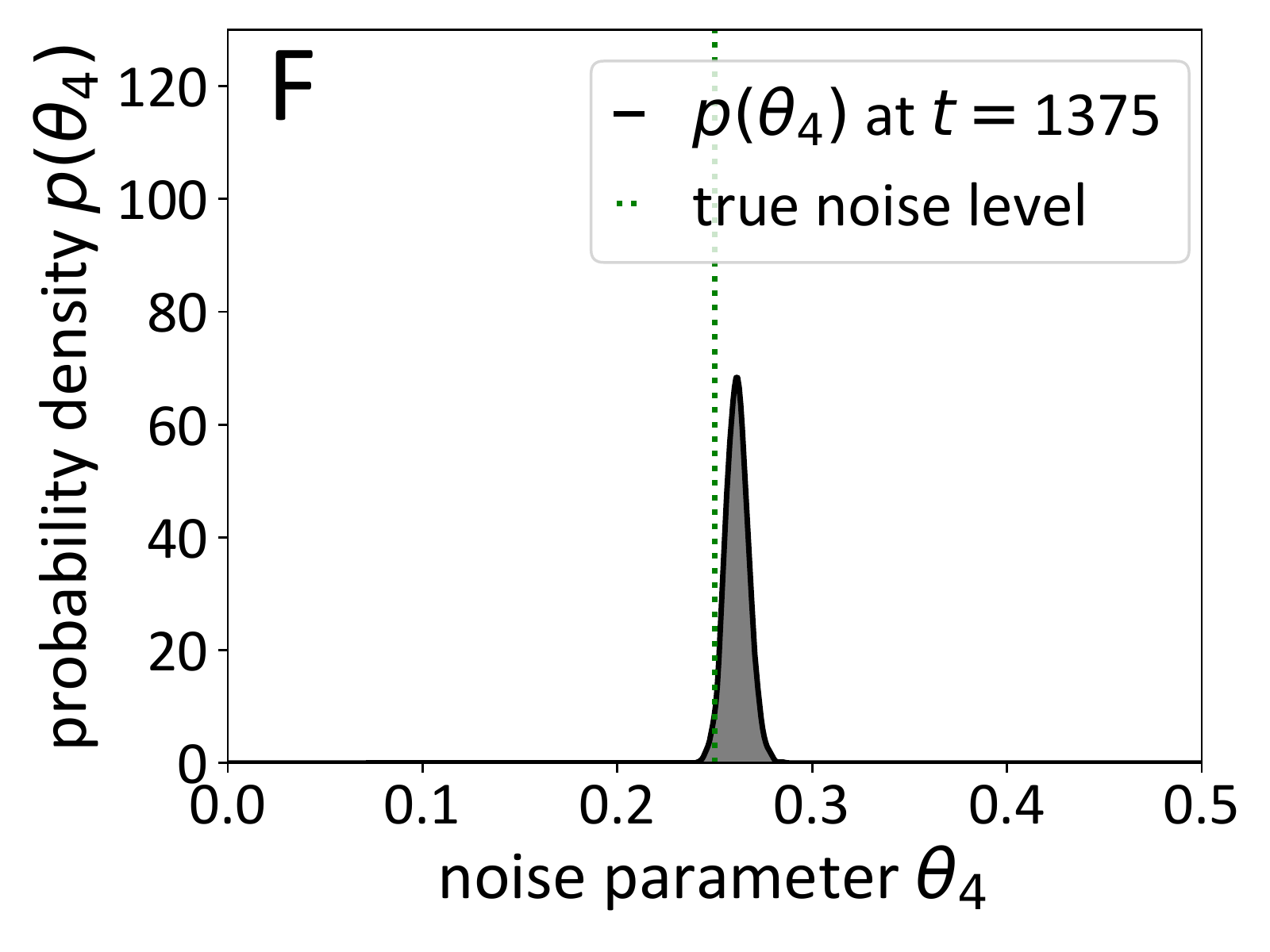}}
\caption[Backstage: Illustration of the numerical procedure.]{Illustration of the key aspects of the numerical procedure. (A) A realisation of the fold model governed by the equations \ref{eq: fold model} is shown. The moving time windows containing $1000$ data points are sketched by the green shaded area. (B) After an initially stable plateau the control parameter $r$ starts to shift linearly. The critical parameter threshold is reached at the red dotted horizontal or vertical lines, respectively. A subjective time estimate for the actual transition is marked by the blue dotted vertical line when the data $x_{\rm fold} < 0.3$ for the first time. The original (blue) and the estimated (orange) potential that underlies the data at that time, is shown in the sub-windows with the data states on the x-axis. The estimation is rather accurate, but the estimated potentials are slightly too narrow. This could be explained due to the noise level that allows only to discover a part of the state space around the fixed point valley of the potential. The Bayesian leading indicator calculation is performed in each window and the probability density of the slope estimates $\hat{\zeta}$ is derived.  (C) The time evolution of the densities is shown. (D-F) The underlying marginal distributions of the parameters $\underline{\theta}$ are shown for the third parameter configurations of the subplots in (B). The pdfs of the parameters $\theta_1$ and $\theta_3$ are relatively defined around zero and suggest a subdominant role of these parameters. The less sharp defined pdfs of $\theta_0$ and $\theta_2$ as well as the maximum probability density $p(\theta_0 \neq 0) = \max(p(\theta_0))$ would indicate a more dominant role of the latter parameters $\theta_0$ and $\theta_2$. The noise is estimated correctly as indicated by the green dotted vertical line for the true noise level.
}
\label{fig: illustrative explanation}
\end{figure*}

The method is tested on two simple synthetic datasets of which one is given by a SDE with
\begin{linenomath*}
\begin{equation}
\begin{aligned}\label{eq: fold model}
h_{\rm fold}(x) = - r + x + x^3\\
g_{\rm fold} = const. = \sigma ,
\end{aligned}
\end{equation}
\end{linenomath*}
where $r$ is the control parameter. The simulation is performed with the Euler-Maruyama scheme \cite{b:Kloeden1992,b:bronstein} for SDEs. The time range is set to $ t = [0,2000]$. The simulation is performed for a linearly increasing control parameter $r$ from $-15$ to $2$. In consequence, the critical transition due to a fold bifurcation is expected around $r\gtrsim 0$. One realisation with $40000$ data points and a noise level of $\sigma = 0.25$ is shown in subfigure \ref{subfig: fold data} that gives an impression of the numerical estimation procedure. The control parameter $r$, shown in \ref{subfig: r param}, is fixed at $-15$ for the first $15000$ data points, before it increases linearly up to $2$ at time $t = 30000$ where it is fixed again for $10000$ data points. The analysis is done on rolling time windows of size $1000$ as suggested by the green area in \ref{subfig: fold data}. 
Starting with this time series the Bayesian analysis approach is applied to each window in order to get insights in the stability and model parameters of the system. The prior range is chosen adequately wide by the interval $[-50,50]$ for $\theta_{0,1,2,3}$ and $[0,50]$ for the stochastic constant factor $\theta_4$ to guarantee the determination of the posterior distribution by the available data instead of the prior assumptions. The standard deviations of the Gaussian priors for $\theta_{2,3}$ are chosen to be $\sigma_{\theta_i} = \lbrace 4,8\rbrace$, respectively. In each window the parameters $\underline{\theta} = \left( \theta_0, \theta_1, \theta_2 , \theta_3, \theta_4 \equiv \sigma\right)^T$ of the third order model $h(x)\approx \rho (x)$ are estimated as the maximum posterior solutions of the corresponding pdfs. The posterior pdfs $p(\theta_{0,1,2,3,4})$ for the rolling time window ending at time $t=1375$ are illustratively shown in the subfigures \ref{subfig: param01}, \ref{subfig: param23} and \ref{subfig: param4}.These pdfs can be helpful to identify the role of the model parameters: Most of the probability density mass of the parameter $\theta_0$ suggests $\theta_0 >0$, whereas the pdfs of the parameters $\theta_1$ and $\theta_3$ are more defined around zero and could be less significant for the model at hand. The quadratic term $\theta_2$ is not as defined as the latter two what means a considerable probability for $\theta_2 \neq 0$. The noise level pdf $p(\sigma)$ includes the real value of $\sigma = 0.25$ that is marked by the green vertical dotted line. The question is: Do the estimated parameters provide an approximately realistic impression of the dynamical situation? In order to answer this question the true potential and the potential that is estimated via the maximum posterior solutions of the parameters $\underline{\theta}$ are shown as blue and orange lines, respectively, in the sub-windows of subfigure \ref{subfig: r param} at the exemplary time $t= 1375$ and $t=375$, $925$ and $1450$. The blue signed original potential tilts over from the stable fixed point $x^* \approx 2.7$ to a new stable position $x^* < 0$. The estimated orange potential preserves the main features of the original potential and follows the same tilting behaviour for increasing $r$. Generally, it is estimated too narrow as expected due to the limited information that is provided by the time series data in terms of the noise that explores the corresponding phase space. In conclusion, the illustration shows that the estimates are reasonable and can be used to calculate the drift slope $\zeta$ that contains information about the stability and upcoming transitions in the system dynamics. The pdf $p(\zeta )$ and its evolution over time is resolved in \ref{subfig: slope pdfs}. The pdfs at time $t = 375$ and $775$ are in the initial fixed regime and approximately equal as expected. When the parameter $r$ starts its linear upwards trend the slope pdfs approach zero and thus, indicate less stable states and indirectly a parameter change in $r$. Note, that the width of the posterior pdfs $p(\zeta )$ decrease strongly in the vicinity of the critical transition. This could mean that the data are more informative in its vicinity.\\

\begin{figure*}[!htbp]
\subfigure{
\label{subfig: fold slope small}
\includegraphics[width=0.32 \textwidth]{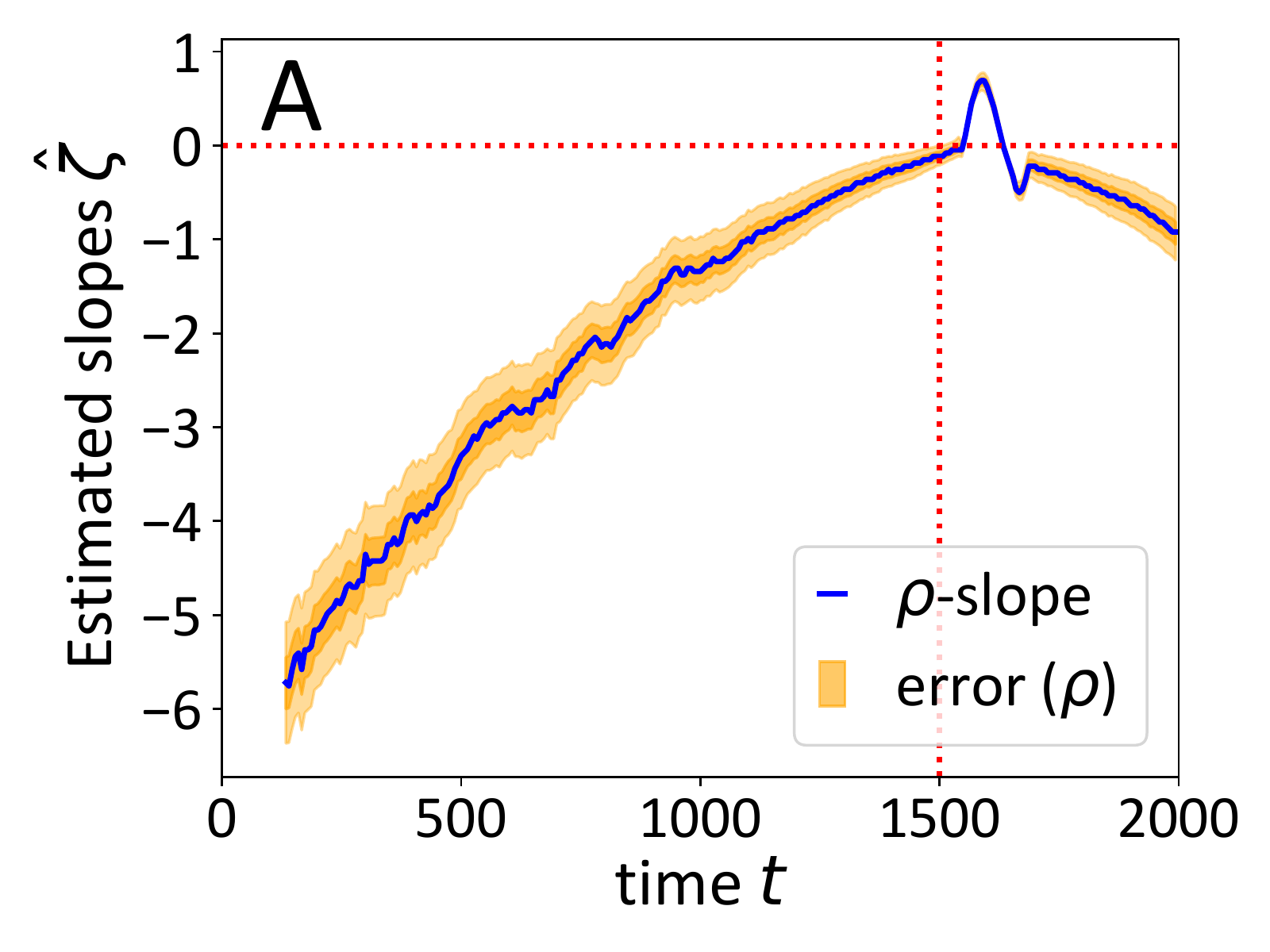}}
\subfigure{
\label{subfig: fold slope medium}
\includegraphics[width=.32 \textwidth]{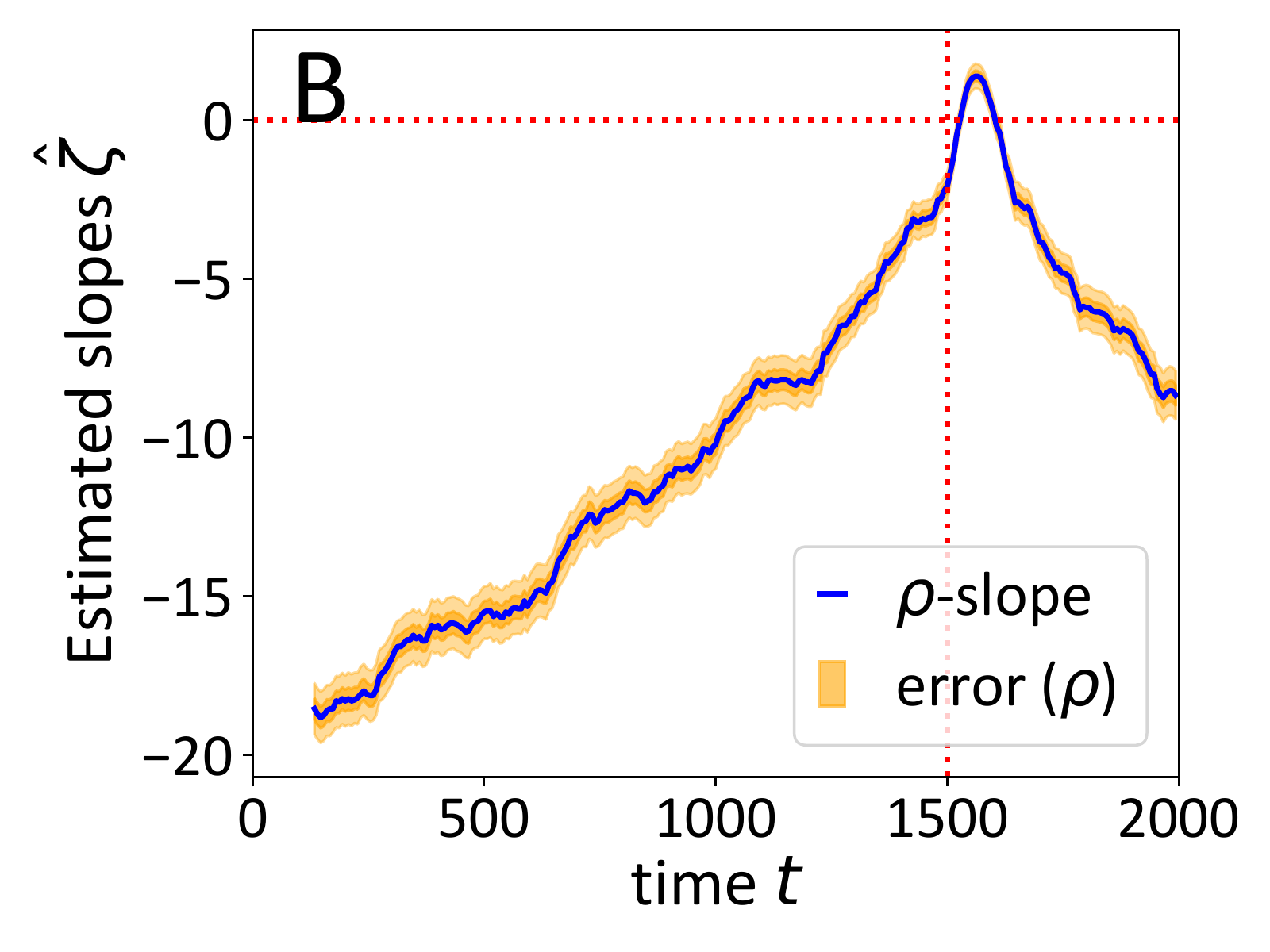}}
\subfigure{
\label{subfig: fold slope high}
\includegraphics[width=.32 \textwidth]{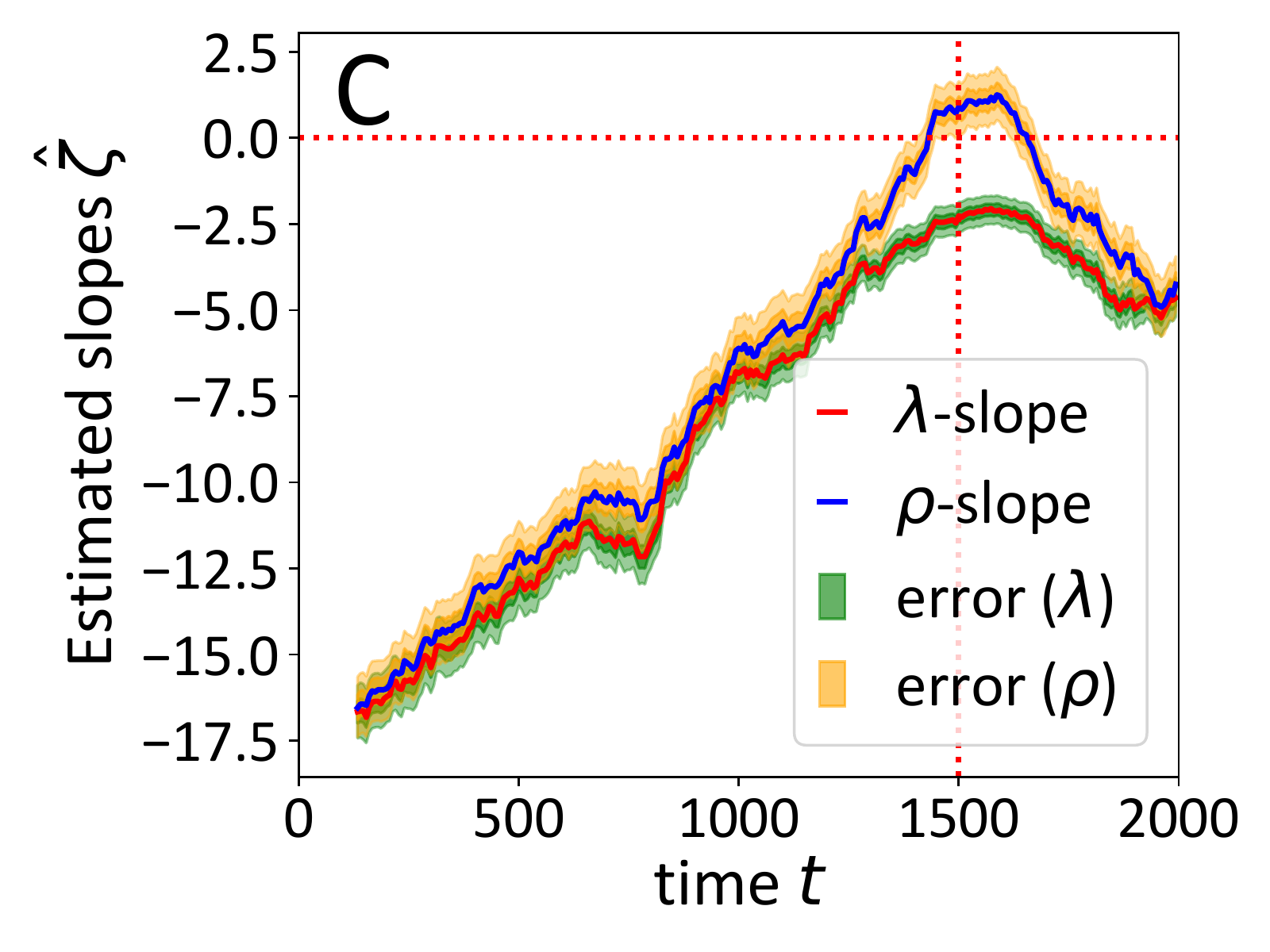}}
\subfigure{
\label{subfig: pitchfork slope small}
\includegraphics[width=0.32 \textwidth]{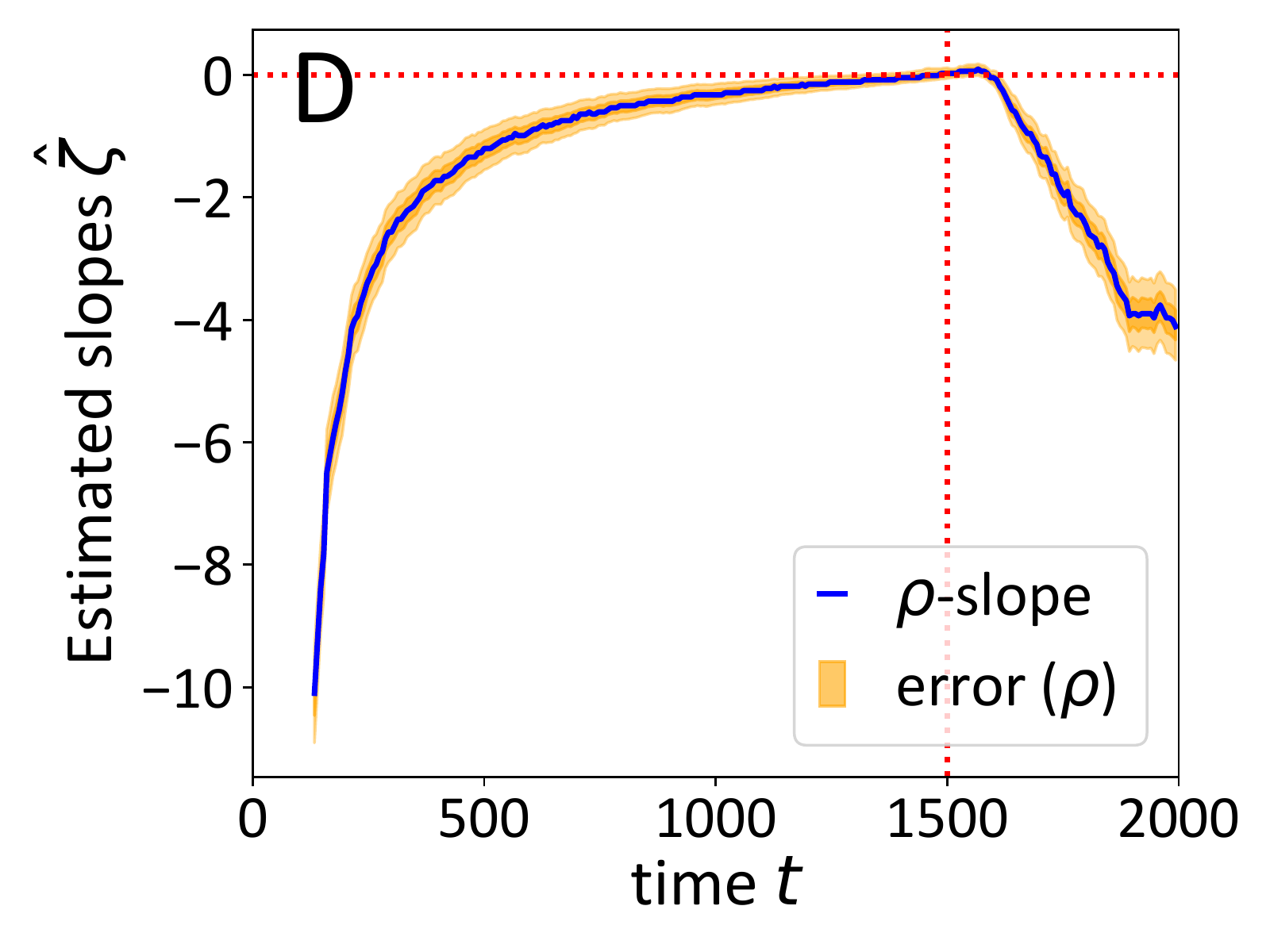}}
\subfigure{
\label{subfig: pitchfork slope medium}
\includegraphics[width=.32 \textwidth]{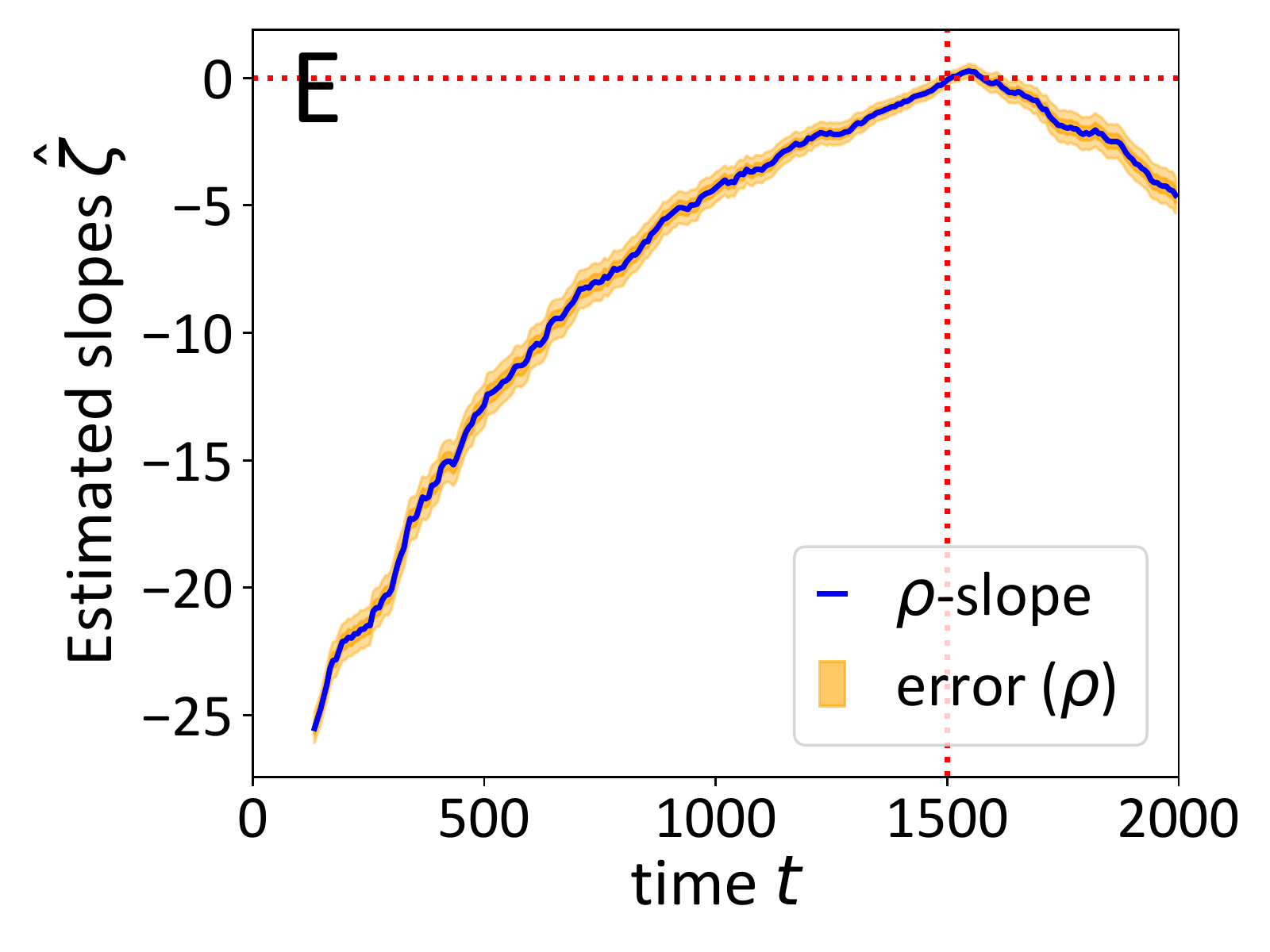}}
\subfigure{
\label{subfig: pitchfork slope high}
\includegraphics[width=.32 \textwidth]{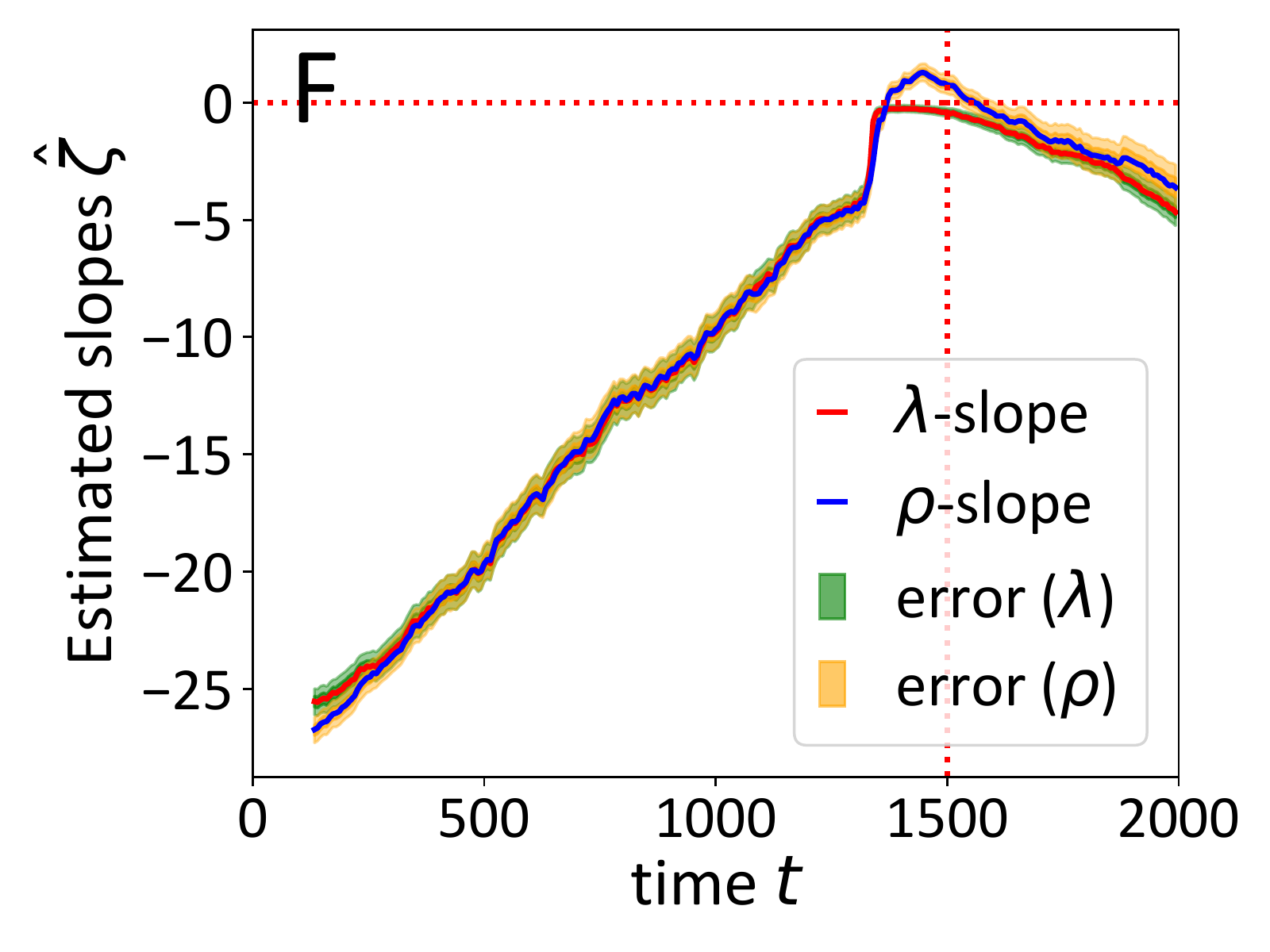}}
\caption[Case studies: Drift slope trends for fold and pitchfork models.]{Drift slope trends for fold and pitchfork models. (A-C) The Bayesian drift slope estimation as leading indicator is tested against fold models of $30000$ data points with linear control parameter trend from $r = -15$ to $r=5$ following the equations \ref{eq: fold model}. The window size is $2000$ points. The critical threshold of the control parameter and the indicators are marked by red dotted vertical and horizontal lines, respectively. The error bands are defined by the $\SI{1}{\percent}$ and $\SI{16}{\percent}$- and $\SI{84}{\percent}$ and $\SI{99}{\percent}$-percentiles of the slope probability density. For small and medium noise levels as $\sigma = 0.05$ in (A) and $\sigma = 0.25$ in (B) the linear and the polynom estimator provide almost identical results apart from the transition region in which the linear slope estimates reach the critical threshold $r_{\rm crit}$, but do not cross it as the polynom counterparts. The advantage of the polynom approximations is evident in the very noisy situation with $\sigma = 2.35$ in (C): The linear estimator does not reach the threshold whereas the polynom estimator marks the flickering region better and moreover, fits well the actual time where the critical threshold $r_{\rm crit}$ is reached. (D-F) A similar result is provided by the estimates of the pitchfork model, governed by the equations \ref{eq: pitchfork}. The trends are almost identical for the two cases with smaller noise ($\sigma = 0.05$, $\sigma = 0.25$), but in the case of high noise $\sigma = 0.75$ in (F), for which in figure \ref{fig: pitchfork data} an exemplary time series is shown, the polynom estimator indicates the strongly flickering region by a positive value whereas the linear estimator does not cross the threshold zero.}
\label{fig: max slopes}
\end{figure*}

\begin{figure}[]
\includegraphics[width=0.4 \textwidth]{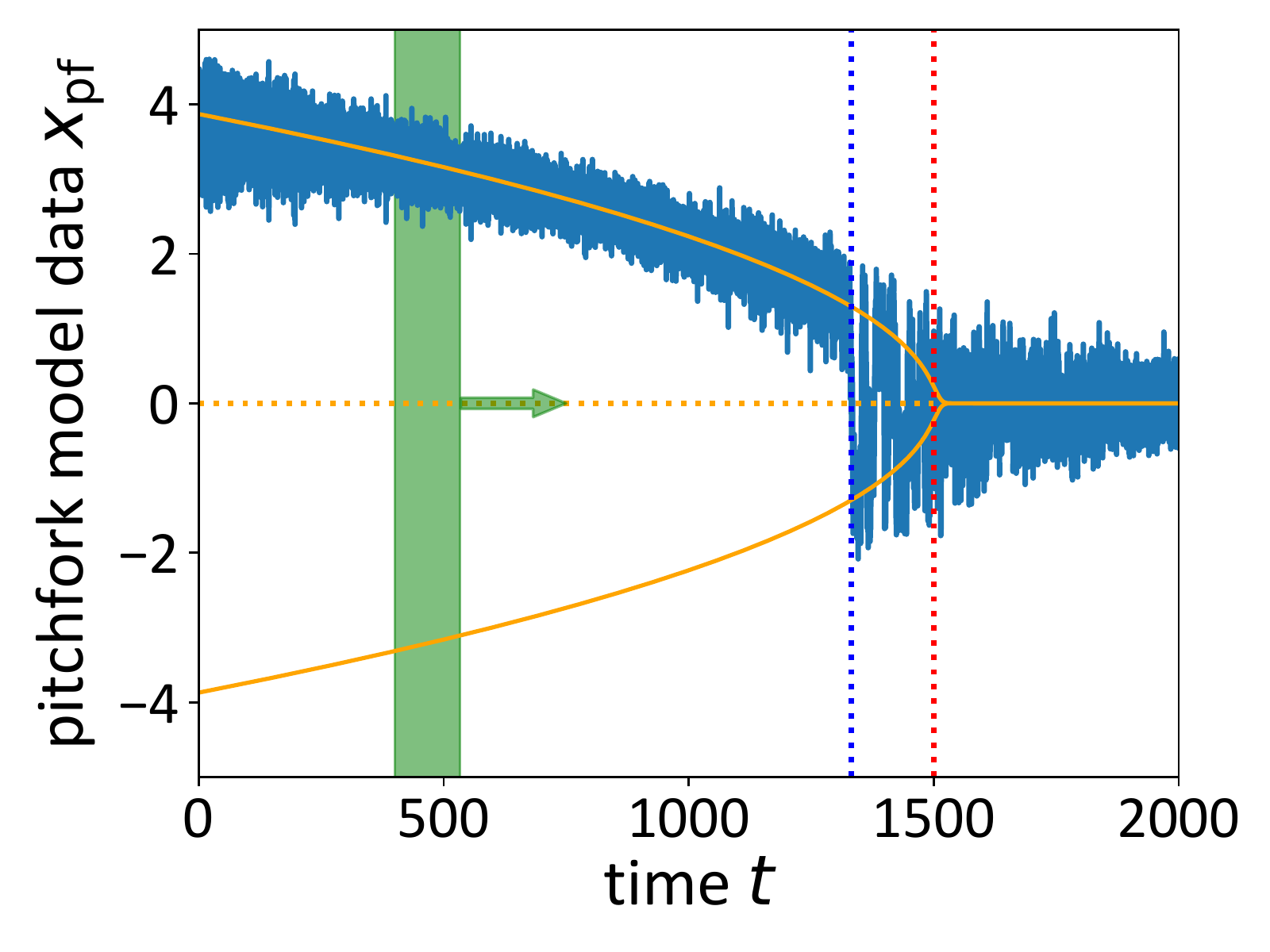}
\caption[An example of the pitchfork model.]{One realisation of the pitchfork model governed by the equations \ref{eq: pitchfork} with the noise level $\sigma = 0.75$ is shown for a linearly shifting control parameter $\nu$. The critical parameter $\nu_{\rm crit}$ is reached at the time $t\approx 1332$ that is marked by the red dotted vertical line. The system exhibits flickering from one to another stable branch starting at the time indicated by the blue dotted vertical line. The stable pitchfork branches are sketched by orange solid lines, the unstable branch is indicated by the orange dotted line. For noise levels $\sigma \geqslant 0.25$ the flickering between the two stable pitchfork branches becomes more and more severe. The dataset is evaluated on time windows of size $2000$. }
\label{fig: pitchfork data}
\vspace{60pt}
\includegraphics[width=0.4\textwidth]{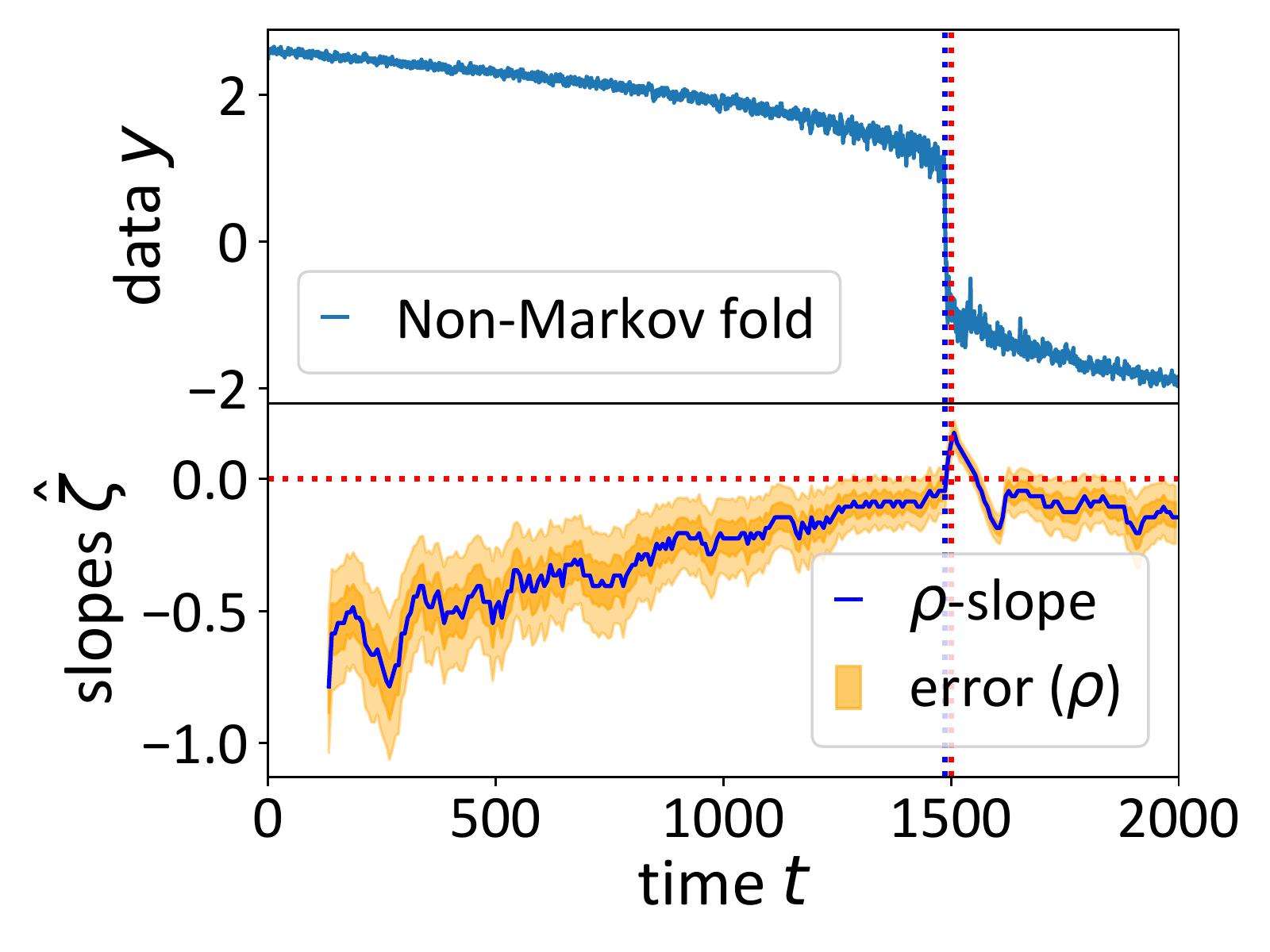}
\caption[Slope estimation for non-Markovian data.]{A realisation of the fold model governed by the equations \ref{eq: non-markovian} for the parameters $q = 0.1$, $c = 0.75$, $d = 5$ and a linearly increasing control parameter $r$ in the range $[-15,5]$ is shown in the upper graph. The critical parameter $r_{\rm crit}$ is reached at the time that is marked by the red dotted vertical line. The blue dotted vertical line illustrates the time at which the dataset is smaller than $0.3$ and does not return. It marks approximately the beginning of the critical transition. The dataset is evaluated in time windows of size $2000$ with the Bayesian slope estimation method introduced in the paper. The resulting slope trend is shown in the lower plot. The results suggest a shift in the absolute values of the estimation, whereas the overall trend that indicates the changing stability is not affected by the non-Markovian feature of the time series.}
\label{fig: non-markovian}
\end{figure}

Thus, the tipping point is correctly indicated by the trend of the MCMC estimate of the drift slope $\zeta$. Figure \ref{fig: max slopes} shows the time evolution of the maximum posterior slope estimates including the corresponding credibility intervals for the standard deviations $\sigma = 0.05$, $\sigma = 0.75$ and $\sigma = 2.35$ of the noise. The results estimated with the linear model $\lambda$ and the polynom model $\rho$ for small and moderate noise contributions as $\sigma = 0.05$ and $\sigma = 0.75$ are almost the same. Therefore, only the polynom estimates are shown in \ref{subfig: fold slope small} and \ref{subfig: fold slope medium}. Only around the actual bifurcation point the results differ slightly, i.d. the linear order model $\lambda$ saturates at $\zeta \approx 0$ instead of the threshold crossing seen for the polynom model $\rho$. The difference in the precision and trend quality is evident for higher noise influence as compared in subfigure \ref{subfig: fold slope high}: The linear model $\lambda$ cannot resolve the critical threshold crossing at the bifurcation point whereas the polynom model is able to handle the noisy situation. Note, that for moderate noise like the scenario investigated in subfigure \ref{subfig: fold slope medium} the slope estimates $\hat{\zeta}$ do not reach the critical threshold, but contain a small jump. In contrast to the scenario with $\sigma = 2.35$ there is no pronounced flickering and the phase space is not explored by a sufficient noise level to anticipate the critical transition exactly.\\
The investigations are proceeded by applying the Bayesian leading indicator approach to another model in order to control its robustness regarding bifurcation types and systems. To this end, a simple model that exhibits a pitchfork bifurcation is computed via a SDE with 
\begin{linenomath*}
\begin{equation}
\begin{aligned}\label{eq: pitchfork}
h_{\rm pf}(x) &=  - \nu \cdot x - x^3 \\
g_{\rm pf} &= const. = \sigma
\end{aligned}
\end{equation}
\end{linenomath*}
with the control parameter $\nu$ that changes linearly in time from $-15$ to $5$ during the simulation with the Euler-Maruyama scheme. The parameters are chosen analogously to the fold model parameters in \ref{fig: max slopes}. The two valleys of the underlying double well potential converge to the unstable branch due to the linear shift in $\nu$ and a single stable branch around zero remains for $\nu \geqslant 0$. The method is tested against this model for different noise levels of which the cases with standard deviation $\sigma = 0.05$, $\sigma = 0.25$ and $\sigma = 0.75$ are shown in this paper. An exemplary dataset with $\sigma = 0.75$ is presented in figure \ref{fig: pitchfork data}. The orange branches indicate stable fixed point positions and the orange dotted horizontal line an unstable fixed point of the typical pitchfork. The data follows the upper branch up to the time $t_{\rm flicker}$ signed by the blue dotted vertical line where flickering occurs. The noise level $\sigma = 0.75$ lies widely in the flickering regime that starts around $\sigma \approx 0.25$ in the simulations' setup. The simulated datasets with standard deviation $\sigma = 0.05$ and $\sigma = 0.25$ follow a similar behaviour but do not show pronounced flickering near the critical threshold $\nu_{\rm crit}$ that is reached at the red dotted vertical line. The Bayesian stability analysis approach turns out to be able to anticipate the ongoing destabilization processes in the pitchfork datasets. The linear as well as the polynom model, $\lambda (x)$ and $\rho (x)$, respectively, yield very similar results for the smaller noise cases as presented in the subfigures \ref{subfig: pitchfork slope small} and \ref{subfig: pitchfork slope medium}. For that reason only the polynom estimations are shown. The advantage of the polynom model $\rho (x)$ lies again in the better handling of highly noisy situations as visible in \ref{subfig: fold slope high}. Of course, the flickering caused by the high noise is not anticipated by none of the estimators. There is a jump in the indicator trend when the flickering starts around the blue dotted vertical line. Nevertheless, the flickering regime up to the critical transition at the critical threshold $\nu_{\rm crit}$ is reproduced rather correctly by the slope estimates $\hat{\zeta} > 0$ with the polynom estimator $\rho (x)$, whereas the linear estimator $\lambda (x)$ does not cross the stability threshold zero.\\
The robustness against high noise is a big problem of known leading indicator candidates. To this end, the performance of the introduced slope estimate $\hat{\zeta}$ and already known leading indicator candidates is compared in the following subsection \ref{subsec: comparison}.
\subsection{\label{subsec: comparison} Robustness against noise}
In many real world applications an analyst has to deal with highly noisy measurement environments. Therefore, the ability of a leading indicator candidate to provide reliable anticipation results is considered in a comparative way: The commonly discussed indicators autocorrelation at lag-1 (AR1), standard deviation (std $\hat{\sigma}$), skewness $\Gamma$ and kurtosis $\omega$ are compared with the introduced drift slope indicator $\zeta$ for the fold model of equation \ref{eq: fold model} in a wide range of noise levels $\sigma=0.05$, $0.25$, $2.35$, $7$, $10$, $12.5$ and $15$. Thus, the noise performance scan incorporates lots higher noise levels than the proposed real noise level of an ecological system around $\sigma = 4.5$ \cite{a:Perretti2012}. Nevertheless, a highly resolved time series and the gradient of a shifting control parameter remain important restrictions to apply the above mentioned indicators successfully.\\
In order to avoid the influence of the window size in a dynamic approach and to include a comprehensive set of noise realisations the procedure is as follows: First, 240 time series of $20000$ data points are simulated over the time interval $[0,2000]$ for each investigated noise level each of which with an equidistantly changed control parameter $r$ over the range $[-15,1]$. This avoids the influence of rolling window size and dynamic changes taking place in a finite window. Second, the test set of leading indicators is applied to the time series generated that way over the whole $20000$ data points to incorporate the noise dynamics. In that way, 240 leading indicator results depending on the control parameter $r$ are computed for each noise level and truncated at the time where $r\geq 0$. Third, these indicator series $\underline{\mathcal{I}}(r)$ are checked with a Bayesian model comparison \cite{b:linden2014} whether they exhibit a linear gradient (model 1 $\mathcal{M}_1$) or are just constant (model 2 $\mathcal{M}_2$). The computed Bayes factors ($\rm BF$) are defined as the fraction of the model evidences $p(\underline{\mathcal{I}}|M_{1,2})$ where $\rm BF_{12}$ indicates the evidence of $\mathcal{M}_1$ over $\mathcal{M}_2$. The Bayes factors are calculated with a standard Gaussian prior for the intercept with means defined by the first data value $\mathcal{I}_0$. A uniform prior in the range $[0, 1.5\cdot (\max(\underline{\mathcal{I}}) - \min(\underline{\mathcal{I}})]$ or $[- 1.5\cdot (\max(\underline{\mathcal{I}}) - \min(\underline{\mathcal{I}})],0]$ for some decreasing skewness cases is chosen for the slope, respectively, and the noise is included with a uniform noise prior in the range $[\log(0.5), \log(5)]$. The convergence of the MC evaluations is guaranteed by $10^7$ draws of prior models in each Bayes factor calculation.\\
The results are summarized in table \ref{table: Bayes factor}. A Bayes factor $\rm BF_{12,21} > 10$ is assumed to be a significant sign of preferring $\mathcal{M}_1$ to $\mathcal{M}_2$ or vice versa.  Following this threshold clearly pronounced gradients are highlighted in green, clearly identified constants in orange and insignificant values in grey in table \ref{table: Bayes factor}. The slope indicator $\zeta$ exhibits significant positive trends in a realistic and wide range of noise levels $\sigma = [0.05,\ 12.5]$, whereas the often mentioned AR1 and std $\hat{\sigma}$ are not as reliable as one may expect. The AR1 is significant only in the small noise regime $\sigma = [0.05,2.35]$ and sensitive to high time series resolutions. A too narrow sampling causes a high autocorrelation at lag-1 and thus, weakens the overall trend. This could be avoided by lower sampling or calculation of  autocorrelation at a higher time lag. The std $\hat{\sigma}$ is only significant for $2.35 \lesssim \sigma \lesssim 7$ and sensitive to a low resolution of the time series. The skewness $\Gamma$ provides good performance results in the high noise regime $\sigma \gtrsim 2.35$ without any exception.  For smaller noise levels the skewness $\Gamma$ is insignificant because of the absence of flickering. The kurtosis $\omega$ has a very bad performance regarding the Bayes factor results. For low noise levels it is insignificant, for a noise around $\sigma \approx 2.35$ positive and for all tested higher noise cases it is described by a constant model which would erroneously indicate no transition.\\

\begin{table*}
\ra{1.4}
    \centering
    \begin{NiceTabular}{@{}lrrrrrrr}[colortbl-like]\toprule[1.3pt]
    noise level &  $\sigma = 0.05$ & $\sigma = 0.25$ & $\sigma = 2.35$ & $\sigma = 7$ & $\sigma = 10$ & $\sigma = 12.5$ & $\sigma = 15$ \\ \cmidrule{1 - 8} 
    indicator $\mathcal{I}$
    & \begin{tabular}{@{}r@{}} $\rm BF_{12}$ \\ $\rm BF_{21}$ \end{tabular} 
    & \begin{tabular}{@{}r@{}} $\rm BF_{12}$ \\ $\rm BF_{21}$ \end{tabular} 
    & \begin{tabular}{@{}r@{}} $\rm BF_{12}$ \\ $\rm BF_{21}$ \end{tabular}
    & \begin{tabular}{@{}r@{}} $\rm BF_{12}$ \\ $\rm BF_{21}$ \end{tabular} 
    & \begin{tabular}{@{}r@{}} $\rm BF_{12}$ \\ $\rm BF_{21}$ \end{tabular} 
    & \begin{tabular}{@{}r@{}} $\rm BF_{12}$ \\ $\rm BF_{21}$ \end{tabular} 
    & \begin{tabular}{@{}r@{}} $\rm BF_{12}$ \\ $\rm BF_{21}$ \end{tabular} \\ 
    \midrule
    slope $\zeta$
    & \cellcolor{green}\begin{tabular}{@{}r@{}} $1.1 \cdot 10^{92}$ \\ $9.6 \cdot 10^{-93}$ \end{tabular}
    & \cellcolor{green}\begin{tabular}{@{}r@{}} $6.2 \cdot 10^{86}$ \\ $1.6 \cdot 10^{-87}$ \end{tabular}
    & \cellcolor{green}\begin{tabular}{@{}r@{}} $2.4 \cdot 10^{53}$ \\ $4.1 \cdot 10^{-54}$ \end{tabular}
    & \cellcolor{green}\begin{tabular}{@{}r@{}} $2.1 \cdot 10^{103}$ \\ $4.8 \cdot 10^{-104}$ \end{tabular}
    & \cellcolor{green}\begin{tabular}{@{}r@{}} $1.5 \cdot 10^{20}$ \\ $6.6 \cdot 10^{-21}$ \end{tabular}
    & \cellcolor{green}\begin{tabular}{@{}r@{}} $1.9 \cdot 10^{3}$ \\ $5.1 \cdot 10^{-4}$ \end{tabular}
    & \cellcolor{gray}\begin{tabular}{@{}r@{}} $1.9$ \\ $0.5$ \end{tabular} \\
    AR1
    & \cellcolor{green}\begin{tabular}{@{}r@{}} $5.4 \cdot 10^{6}$ \\ $1.9 \cdot 10^{-7}$ \end{tabular}
    & \cellcolor{green}\begin{tabular}{@{}r@{}} $1.5 \cdot 10^{7}$ \\ $6.9 \cdot 10^{-8}$ \end{tabular}
    & \cellcolor{green}\begin{tabular}{@{}r@{}} $8.1 \cdot 10^{10}$ \\ $1.2 \cdot 10^{-11}$ \end{tabular}
    & \cellcolor{gray}\begin{tabular}{@{}r@{}} $1.9$ \\ $0.5$ \end{tabular}
    & \cellcolor{gray}\begin{tabular}{@{}r@{}} $1.0$ \\ $1.0$ \end{tabular}
    & \cellcolor{gray}\begin{tabular}{@{}r@{}} $1.0$ \\ $1.0$ \end{tabular}
    & \cellcolor{gray}\begin{tabular}{@{}r@{}} $1.0$ \\ $1.0$ \end{tabular} \\
    std $\hat{\sigma}$
    & \cellcolor{gray}\begin{tabular}{@{}r@{}} $1.0$ \\ $1.0$ \end{tabular}
    & \cellcolor{gray}\begin{tabular}{@{}r@{}} $0.26$ \\ $3.8$ \end{tabular}
    & \cellcolor{green}\begin{tabular}{@{}r@{}} $2.7 \cdot 10^{12}$ \\ $3.7 \cdot 10^{-13}$ \end{tabular}
    & \cellcolor{green}\begin{tabular}{@{}r@{}} $200$ \\ $4.9\cdot 10^{-3}$ \end{tabular}
    & \cellcolor{gray}\begin{tabular}{@{}r@{}} $2.6$ \\ $0.4$ \end{tabular}
    & \cellcolor{gray}\begin{tabular}{@{}r@{}} $1.2$ \\ $0.8$ \end{tabular}
    & \cellcolor{gray}\begin{tabular}{@{}r@{}} $1.1$ \\ $0.9$ \end{tabular} \\
    skewness $\Gamma$
    & \cellcolor{gray}\begin{tabular}{@{}r@{}} $1.0$ \\ $1.0$ \end{tabular}
    & \cellcolor{gray}\begin{tabular}{@{}r@{}} $0.18$ \\ $5.4$ \end{tabular}
    & \cellcolor{green}\begin{tabular}{@{}r@{}} $72$ \\ $ 1.4 \cdot 10^{-2}$ \end{tabular}
    & \cellcolor{green}\begin{tabular}{@{}r@{}} $3.3 \cdot 10^{60}$ \\ $3.1 \cdot 10^{-61}$ \end{tabular}
    & \cellcolor{green}\begin{tabular}{@{}r@{}} $1.0 \cdot 10^{71}$ \\ $1.0 \cdot 10^{-72}$ \end{tabular}
    & \cellcolor{green}\begin{tabular}{@{}r@{}} $2.0 \cdot 10^{59}$ \\ $4.9 \cdot 10^{-60}$ \end{tabular}
    & \cellcolor{green}\begin{tabular}{@{}r@{}} $1.6 \cdot 10^{47}$ \\ $6.1 \cdot 10^{-48}$ \end{tabular} \\
    kurtosis $\omega$
    & \cellcolor{gray}\begin{tabular}{@{}r@{}} $1.0$ \\ $1.0$ \end{tabular}
    & \cellcolor{gray}\begin{tabular}{@{}r@{}} $1.9$ \\ $0.53$ \end{tabular}
    & \cellcolor{green}\begin{tabular}{@{}r@{}} $5.8 \cdot 10^{100}$ \\ $1.7 \cdot 10^{-101}$ \end{tabular}
    & \cellcolor{orange}\begin{tabular}{@{}r@{}} $4.6 \cdot 10^{-2}$ \\ $22$ \end{tabular}
    & \cellcolor{orange}\begin{tabular}{@{}r@{}} $3.7 \cdot 10^{-2}$ \\ $27$ \end{tabular}
    & \cellcolor{orange}\begin{tabular}{@{}r@{}} $4.3 \cdot 10^{-2}$ \\ $23$ \end{tabular}
    & \cellcolor{orange}\begin{tabular}{@{}r@{}} $5.9 \cdot 10^{-2}$ \\ $17$ \end{tabular} \\
    \bottomrule
    \end{NiceTabular}
    \caption{The table shows the Bayes factors $\text{BF}_{12,21}$ for the drift slope $\zeta$ and the leading indicator candidates autocorrelation (AR1), standard deviation (std $\hat{\sigma}$), skewness $\Gamma$ and kurtosis $\omega$ regarding a linear gradient (model 1) or a constant (model 2) in a wide range of noise levels $\sigma = [0.05,15]$ incorporating realistic ecological noise $\sigma \approx 4.5$ \cite{a:Perretti2012}. The Bayes factor is defined as significant for $\text{BF}_{12,21} > 10$. A pronounced positive trend as a hint for an uprising destabilization is highlighted in green, an insignificant result in grey and preferring the constant model in orange. The AR1 and std $\hat{\sigma}$ are just significant in a small range of noise levels. The drift slope $\zeta$ is robust against noise until $\sigma \lesssim 12.5$ and thus, seems to be a reliable leading indicator candidate. Similar, the skewness $\Gamma$ is a reliable indicator for high noise situations that include the phenomenon of flickering. The kurtosis performs poorly in this analysis and suggests erroneously no transition for noise $\sigma \gtrsim 7$.}
    \label{table: Bayes factor}
\end{table*}

\subsection{\label{subsec: non markovian} Non-Markovian data}
Up to now the noise is assumed to be Gaussian and $\delta$-correlated in time so that the process obeys the Markov properties \cite{b:Kloeden1992, ip:Kloeckner2014}. But it is emphasized that empirical tests suggest that even if the Markov assumption is not fulfilled - i.e. the data is simulated with correlated noise and thus, non-Markovian - the trend of the stability indicator remains the same apart from some shift in its absolute values as discussed in the example in this section. Thus, its interpretation is not necessarily strongly affected.\\
The basic assumption of a Markovian process is not necessarily fulfilled in realistic scenarios. Nevertheless, the heuristic ansatz of the method that is introduced in the paper, does not loose its principal validity: In figure \ref{fig: non-markovian} the method is applied to a realisation of a non-Markovian \cite{Risken96} fold model, i.e.
\begin{linenomath*}
\begin{equation}
\begin{aligned}
\dot{x} = - r + x - x^3 + q \cdot y \\
\dot{y} = - c \cdot y + d \cdot \Gamma(t)
\label{eq: non-markovian}
\end{aligned}
\end{equation}
\end{linenomath*}
with the parameters $q = 0.1$, $c = 0.75$, $d = 5$ and a control parameter $r$ linearly increasing in the range $[-15,5]$. The slope estimates $\hat{\zeta}$ are computed with the method and a polynomial model $\rho$ with constant noise level $\theta_4 = const. = \sigma$. The illustration \ref{fig: non-markovian} suggests the general validity of the method and its ability to determine destabilizing trends even in non-Markovian time series data, whereas the absolute values are affected by a value shift.\\
The case studies on the fold and pitchfork model are completed at this point. In the following subsection \ref{subsec: prediction} the idea of providing a transition time estimate $\hat{t}_{\rm crit}$ with uncertainty bands is proposed.
\subsection{\label{subsec: prediction}Non-parametric anticipation of critical transitions}
\begin{figure*}[]
\subfigure{
\label{subfig: prediction1}
\includegraphics[width=0.32 \textwidth]{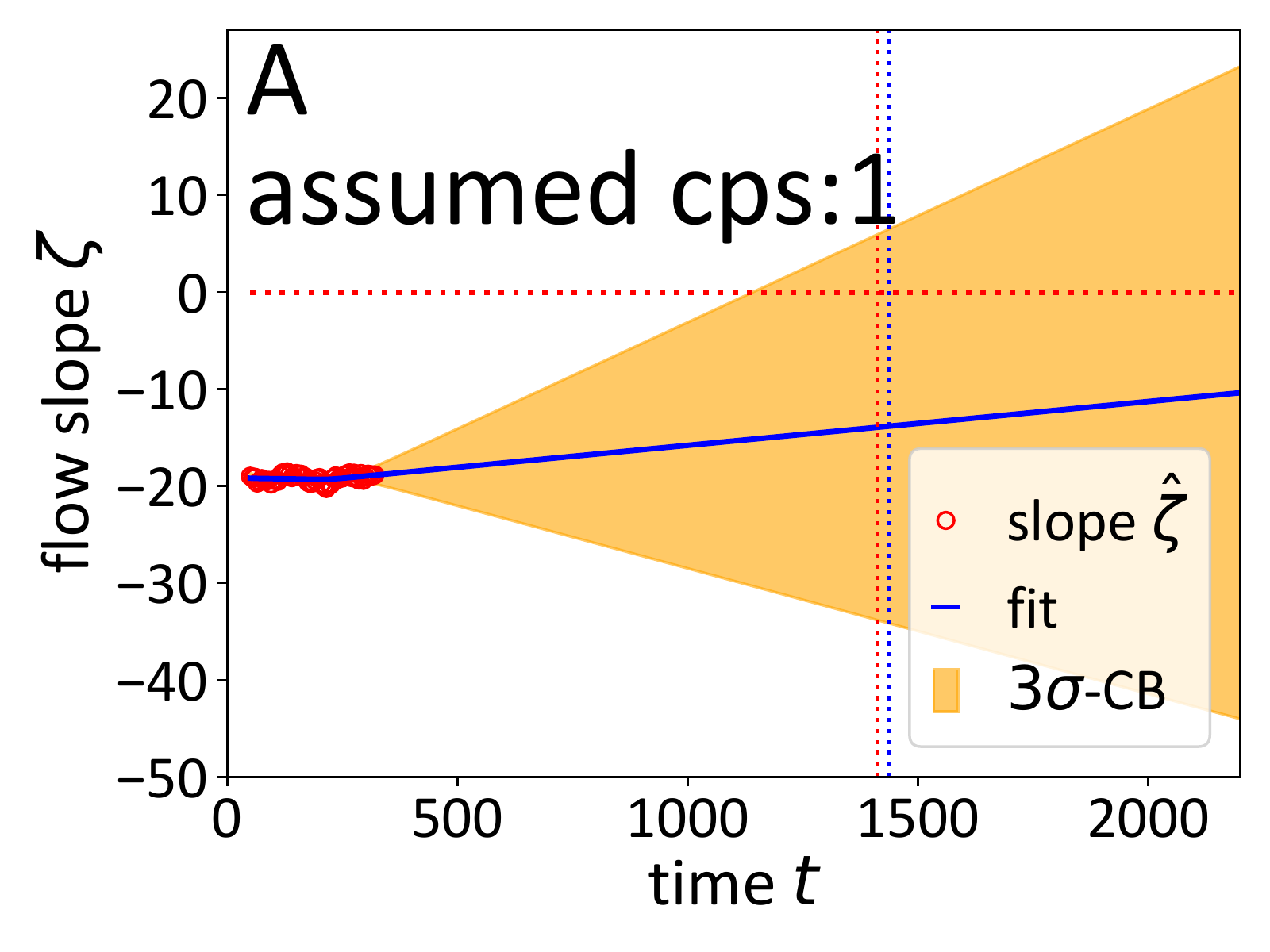}}
\subfigure{
\label{subfig: prediction2}
\includegraphics[width=.32 \textwidth]{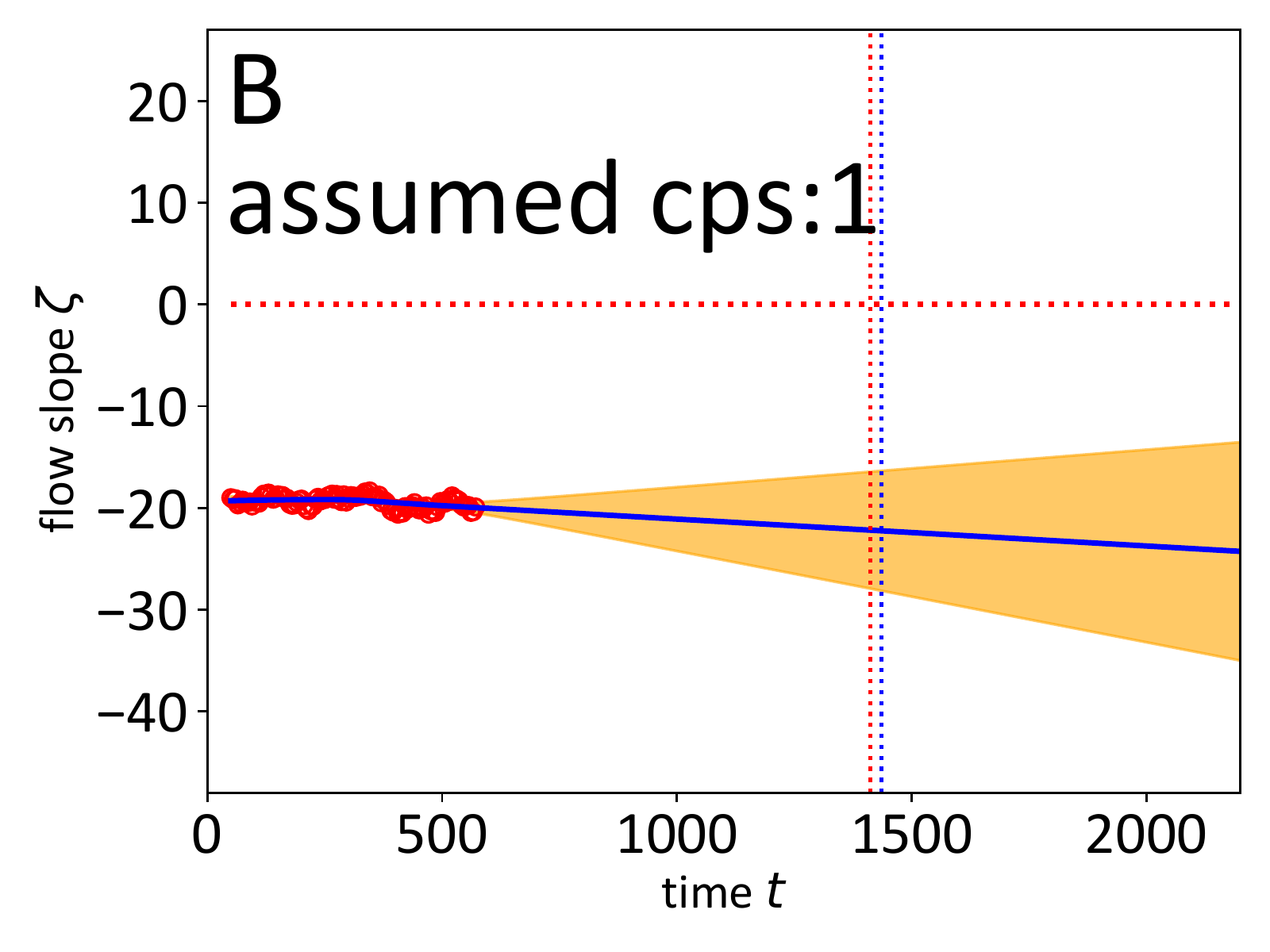}}
\subfigure{
\label{subfig: prediction3}
\includegraphics[width=.32 \textwidth]{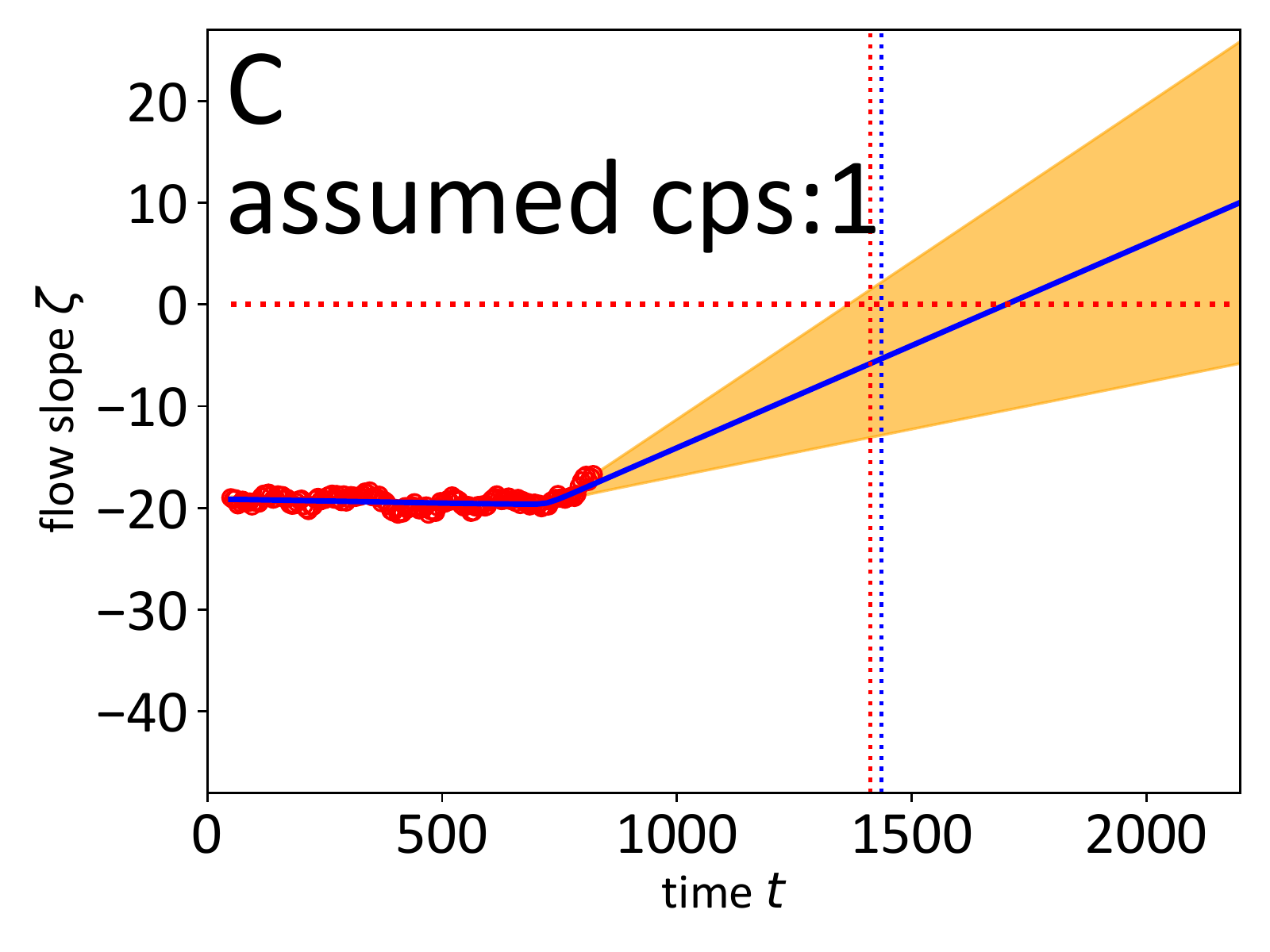}}
\subfigure{
\label{subfig: prediction4}
\includegraphics[width=0.32 \textwidth]{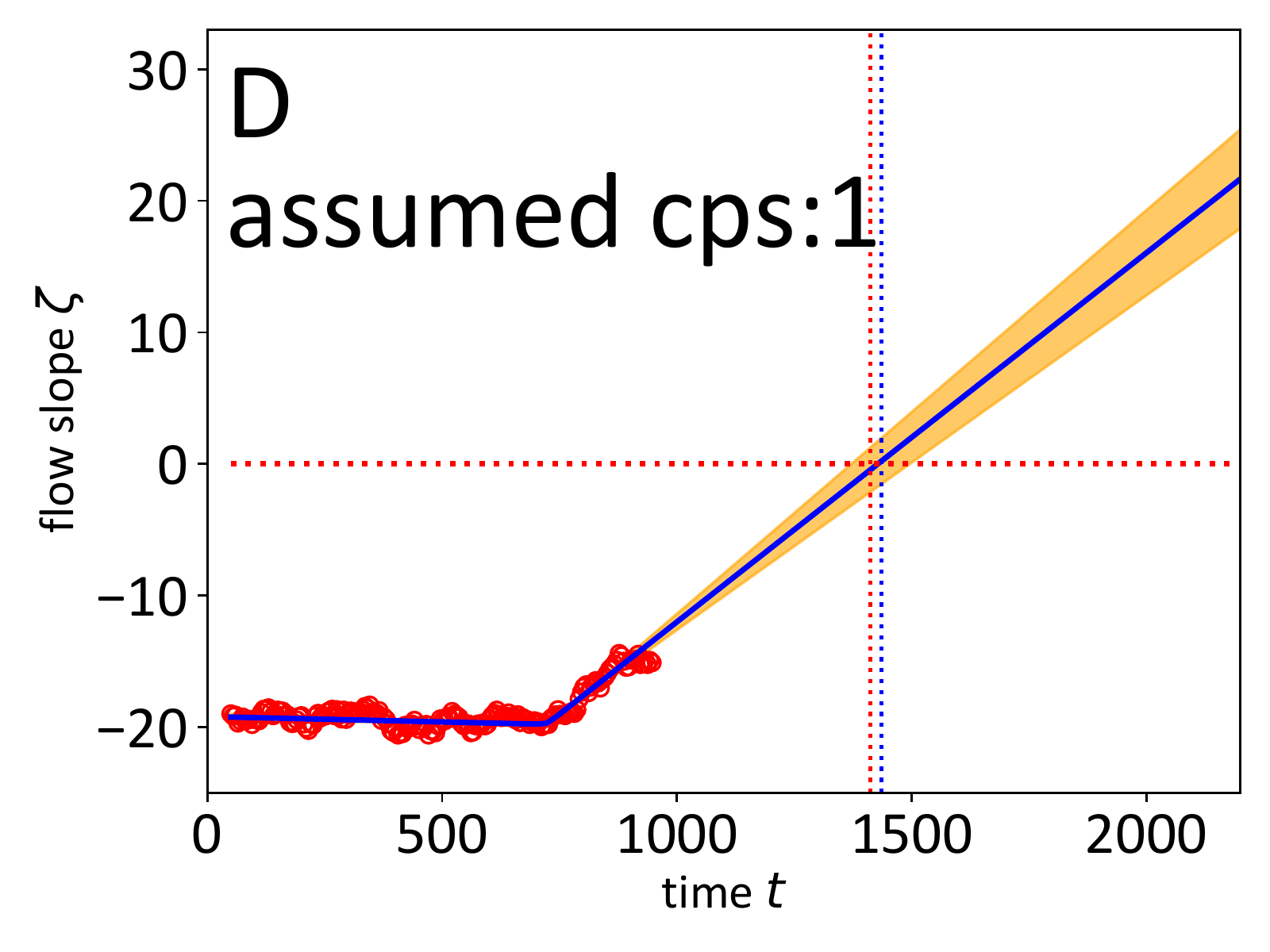}}
\subfigure{
\label{subfig: prediction5}
\includegraphics[width=.32 \textwidth]{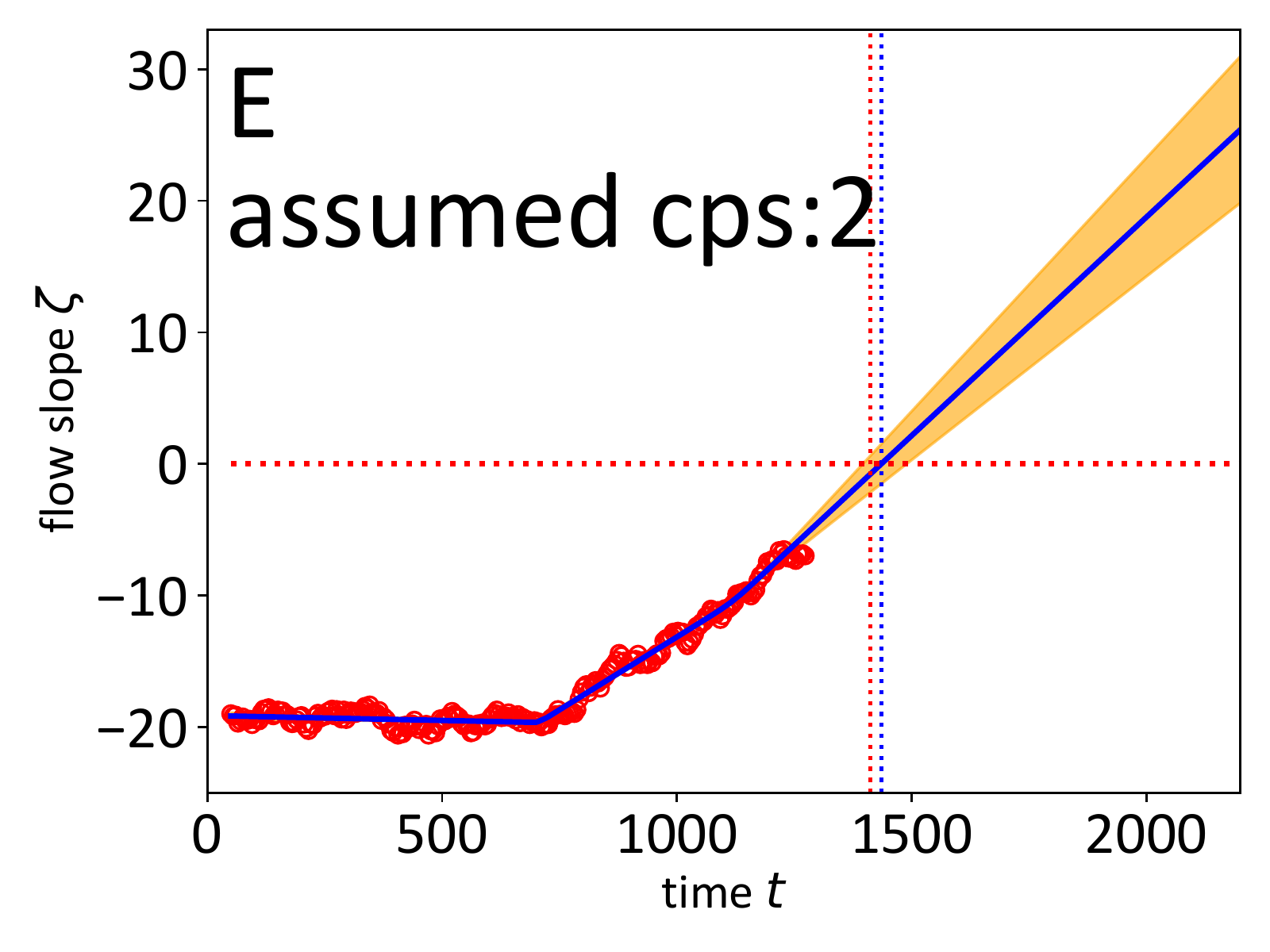}}
\subfigure{
\label{subfig: prediction6}
\includegraphics[width=.32 \textwidth]{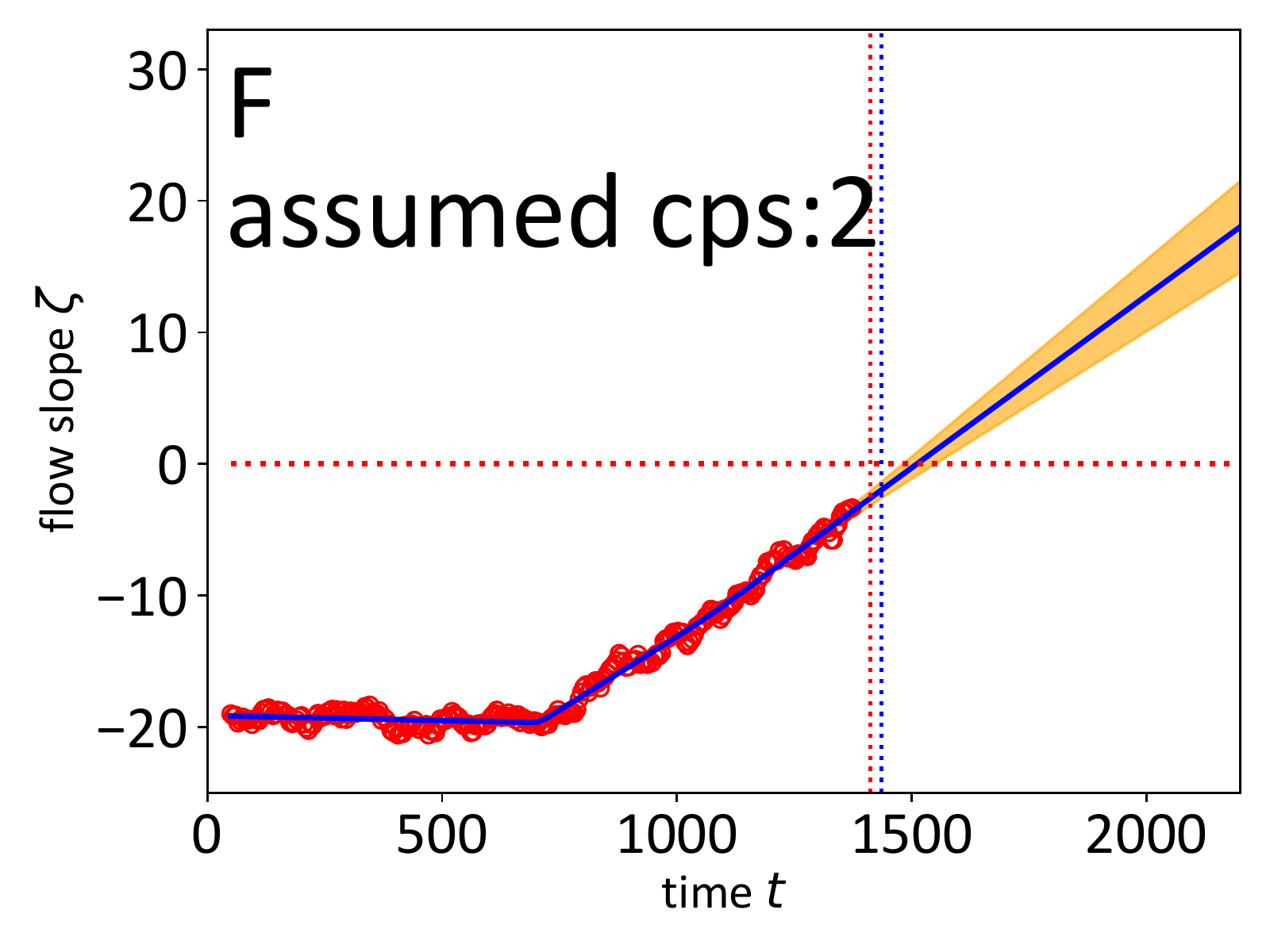}}
\caption[Bayesian segment fits to anticipate critical transitions.]{Six snapshots of the simulated on-line approach to predict the uprising critical transition and the change of the parameter trend. The blue solid lines are the updated linear segment fits based on the slope estimates (red circles) calculated from the dataset shown in \ref{subfig: fold data}. The corresponding uncertainty bands are marked in orange. The red dotted vertical and horizontal lines indicate the time at which the critical control parameter is reached and the drift slope indicator threshold, respectively. A subjective transition time is sketched by the blue dotted vertical line. The drift slope data is successively expanded and the calculations are redone to simulate the on-line scenario. At the beginning one change point is assumed to occur in the data. (A) For a small amount of data a huge uncertainty band proposes a correct interval for the critical transition as visible. (B) More data improve the fit in the stable region and no transition is anticipated in the near future. (C) The change point at time $t\approx 750$ is precisely followed by the fit approach and the extrapolation gives a good forecast of the actual transition time. (D) The same holds true for the next fit and the uncertainty tends to decrease, before a second change point is assumed in (E). The assumption of a second change point increases slightly the uncertainty about the future. (F) In vicinity of the critical transition the fit approach suggests transition time forecasts slightly behind the actual transition. This is an expected result since the rolling window approach incorporates a time delay compared to the true dynamics of the system.
}
\label{fig: prediction}

\subfigure{
\label{subfig: forecast statistics 1}
\includegraphics[width=0.45 \textwidth]{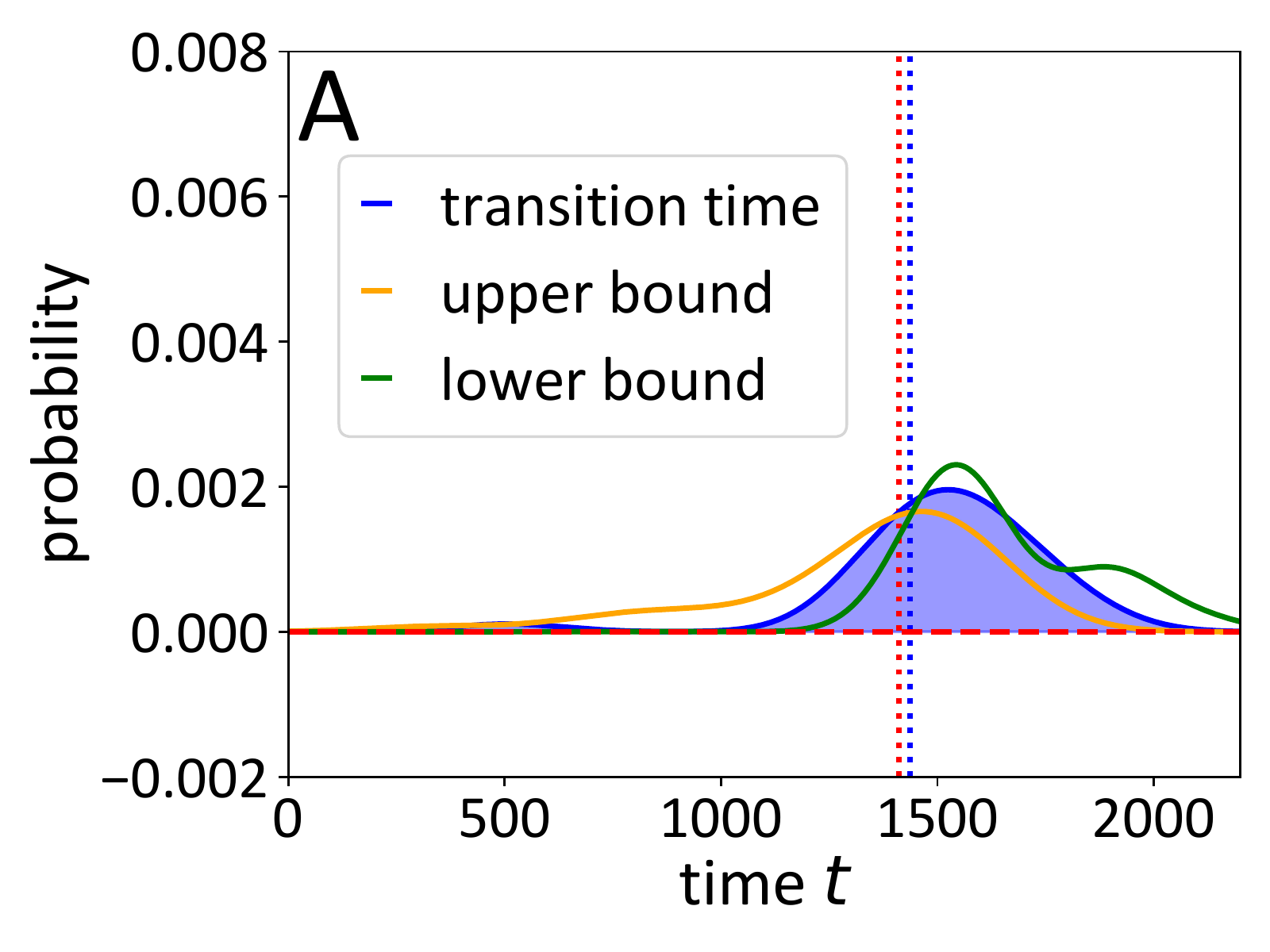}}
\subfigure{
\label{subfig: forecast statistics 2}
\includegraphics[width=.45 \textwidth]{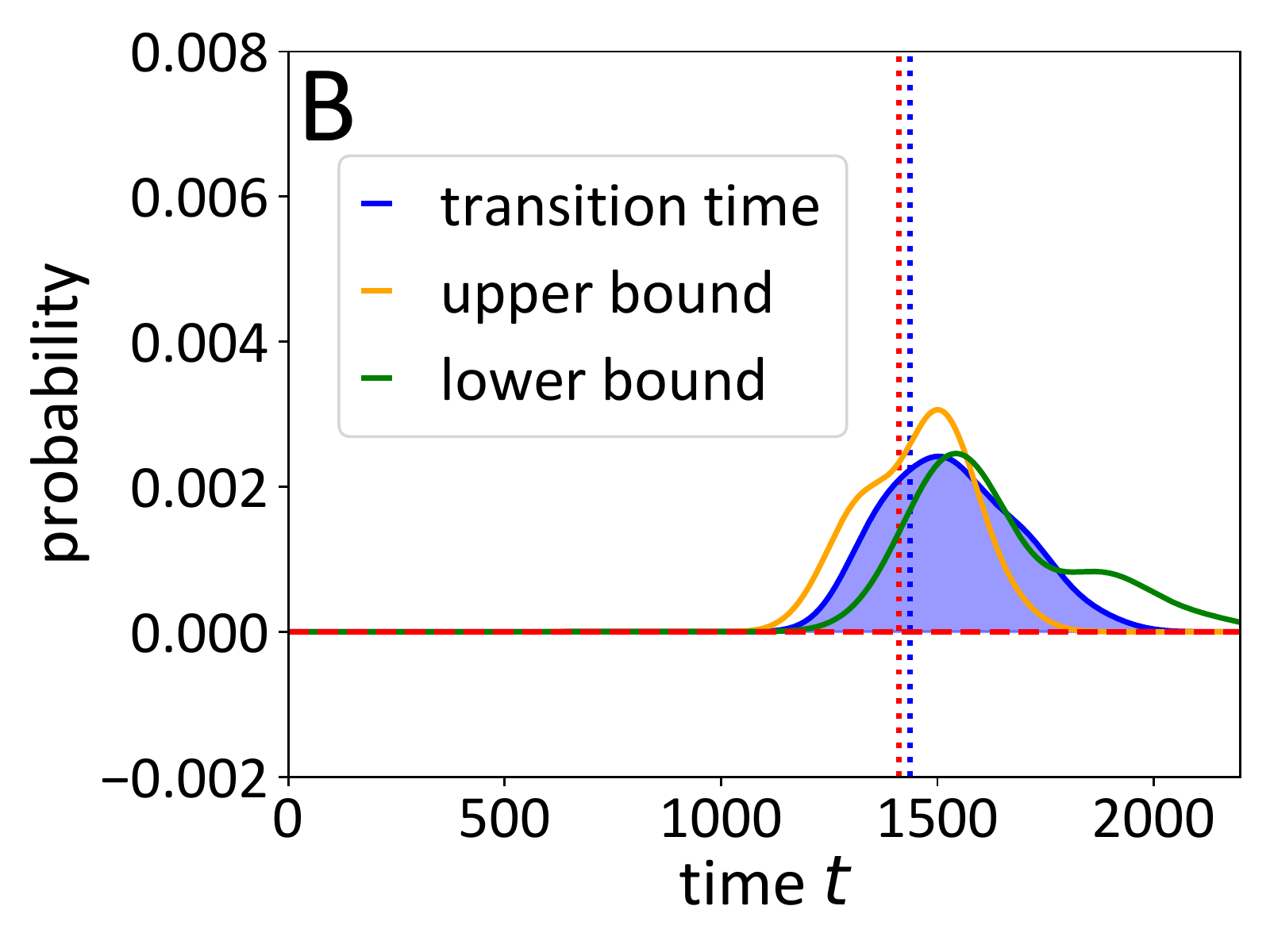}}
\caption[Transition times: forecast statistics]{Statistics of the transition time forecasts and credibility bounds. The red and blue dotted vertical lines indicate the critical parameter threshold and the time at which the fold data decreases below $0.3$ for the first time, respectively. (A) The kernel density estimates of the fitted transition times and the bounds of all performed fits are shown. The distributions suggest a low possibility for a transition prior to the actual transition time (blue dotted line) and increase significantly in its vicinity. (B) The kernel density estimates are computed with the fits performed after time $t = 801.9$. These fits lie behind the first recognized change point. The pdfs are more narrow and thus, the low probability of an early transition vanishes. In both cases (A) and (B) the right transition time is incorporated by the pdfs, although their maxima are behind the actual transition time. This is probably due to a delay introduced by the rolling window approach.}
\label{fig: forecast statistics}
\end{figure*}

The problem of forecasting possible transition times in a reliable manner remains unsolved so far, but the non-parametric fit approach explained in detail in \cite{b:linden2014, a:dose2004} is a useful tool to face the problem. To this end the proposed transition time estimation approach is presented as an example on the slope estimates of the fold dataset that is presented above in \ref{subfig: fold data}. This data exhibits a constant control parameter $r = -15$ up to time $t = 15000$ and then starts to shift linearly up to $r = 2$ at time $t = 30000$. In this paper the on-line application is simulated by successively updating the available slope estimates and repeating the Bayesian fit process in each step. For practical reasons the fitting procedure is started under the assumption of one possible change point. The uprising control parameter trend is detected and well fitted at $t \approx 15000$. After identification of a change point a second possible change point is incorporated in the assumptions to take care of possibly further changing parameter trends. Thus, the assumption of two change points is included for times $t\geqslant 1030$. Snapshots of the extrapolation method are shown in figure \ref{fig: prediction}. The red dotted vertical line marks the time at which the control parameter crosses its critical value. The blue dotted vertical line indicates the actual critical transition defined analogously to subfigure \ref{subfig: fold data}. For the problem at hand it is also reasonable to choose $ 3 \sigma$-credibility intervals for the fit functions to be also aware of critical transitions that could occur much earlier than the most probable transition. In subfigures \ref{subfig: prediction1} and \ref{subfig: prediction2} the control parameter is constant. In both situations no critical transition is anticipated in the near future. Nevertheless, the very small amount of data and the assumption of a possible change point in the segment fit leads to a wide uncertainty range in \ref{subfig: prediction1}, that even proposes a small possibility for a transition in the shown interval of the investigated time series. As the available data of the constant stable regime increases the uncertainty decreases as visible in \ref{subfig: prediction2}. The segment fit detects the uprising destabilization process very precisely as underlined by \ref{subfig: prediction3}. The change point is correctly detected and the extrapolation predicts the correct interval for an uprising transition based on the currently available data around $650$ time points before it occurs. Also in the subfigures \ref{subfig: prediction4} to \ref{subfig: prediction6} the uncertainty range covers the correct interval.\\
The forecast transition times as well as their upper and lower bounds are considered in a statistical analysis over the time interval $t \in [0,2200]$ in order to
\FloatBarrier
\FloatBarrier
get an idea of the reliability of the predictions in that model case.
The results are summarized in figure \ref{fig: forecast statistics}. The probability distributions contain the right transition interval, whereas their maxima are behind the actual transition time. This is an expected result because of the rolling time window method that incorporates a delayed trend in the drift slope calculations. Ne\-ver\-the\-less, the distributions give an idea of the probability of a critical transition for different times in future. After the time $t=801.9$ $\SI{100}{\percent}$ of the $25$ performed fits suggest the actual transition prior to time $t = 2200$ and even $\SI{32.0}{\percent}$ of the individual fits contain the right transition time in their corresponding $3\sigma$ confidence bands.
\section{\label{sec: conclusion} Conclusion and outlook}
In this paper we have introduced a robust method to anticipate critical transitions due to bifurcations and applied it to a system with a fold and a pitchfork bifurcation. The method is based on the hypothesis that the measured data can be described by a parametric nonlinear SDE. The estimation of the coefficients of the SDE allows for a direct access to the linear stability of the system via the slope of the deterministic part of the SDE at the fixed point. By including nonlinear terms into the ansatz the method provides reliable response to destabilization even in the presence of strong noise. Thanks to the Bayesian approach of parameter estimation we have full access to the posterior pdfs of the parameters via MCMC and can consistently define credibility intervals which are important for the interpretation of the results. The upper uncertainty bound of the slope of the SDE's deterministic part yields a reliable limit for the earliest possible transition time with respect to the actual transition time of the two investigated simple models. The inclusion of a Bayesian non-parametric linear segment fit allows for both an on-line determination of change points in the current trend and the estimation of the point where the transition probably will take place in the future based on the current information.\\
The forecast procedure can also be adapted to nonlinear time evolution of the control parameter by introducing more change points. Nevertheless, this increases the computational costs of the method which might be problematic for on-line applications. Therefore, further work has to be done to make the method also feasible for on-line applications with nonlinear trends towards tipping points.
\subsection*{Data availability}
The datasets as well as the simulations and analyis codes are available upon request from the authors under a CC BY-NC-ND 4.0 International license.\\
\subsection*{Acknowledgements}
We thank colleagues and friends for proofreading the manuscript. M. H. thanks the Studienstiftung des deutschen Volkes for a scholarship including financial support.

\bibliography{main}

\end{document}